\documentclass[a4paper,useAMS,usenatbib]{mn2e}

\usepackage{graphicx}
\usepackage[T1]{fontenc}
\usepackage{amsmath}
\usepackage{aas_macros}
\usepackage{times}
\usepackage{ amssymb }
\usepackage{color}
\usepackage[total={17.8cm,24.0cm},centering]{geometry}

\title[Dissecting galactic bulges in space and time I]
{Dissecting galactic bulges in space and time I: \newline the importance of early formation scenarios vs. secular evolution}

\author[Seidel et al.]{M.~K.~Seidel$^{1,2}$\thanks{E-mail: mseidel@iac.es},
R.~Cacho$^{3}$, T.~Ruiz-Lara$^{4,5}$, J.~Falc\'{o}n-Barroso$^{1,2}$,
I.~P\'erez$^{4,5}$, 
\newauthor
P. S\'anchez-Bl\'azquez$^{6}$, F.~P.~A.~Vogt$^{7}$, M.~Ness$^{8}$, K.~Freeman$^{7}$
and S.~Aniyan$^{7,9}$\\
$^{1}$Instituto de Astrof\'isica de Canarias, E-38200 La Laguna, Tenerife,
Spain\\
$^{2}$Departamento de Astrof\'isica, Universidad de La Laguna, E-38205 La
Laguna, Tenerife, Spain\\
$^{3}$Dpto. de Astrof\'isica y CC. de la Atm\'osfera, Universidad Complutense de
Madrid, Spain\\
$^{4}$Dpto. de F\'isica Te\'orica y del Cosmos, Universidad de Granada,
Facultad de Ciencias (Edificio Mecenas), 18071 Granada, Spain\\
$^{5}$Instituto Universitario Carlos I de F\'isica Te\'orica y Computacional,
Universidad de Granada, 18071 Granada, Spain\\
$^{6}$Dpto. de F\'isica Te\'orica, Universidad Aut\'onoma de Madrid, E-28049
Cantoblanco, Spain\\
$^{7}$Research School of Astronomy and Astrophysics, Australian National University, Canberra, ACT 2611, Australia\\
$^{8}$Max-Planck-Institut f\"ur Astronomie, K\"onigstuhl 17, D-69117 Heidelberg, Germany \\
$^{9}$European Southern Observatory, Karl-Schwarzschild-Str. 2, 85748 Garching, Germany}

\begin{document}

\pagerange{\pageref{firstpage}--\pageref{lastpage}} \pubyear{2002}
\maketitle
\label{firstpage}

%=============================================================================
\begin{abstract}
The details of bulge formation via collapse, mergers, secular processes or their interplay remain unresolved.
To start answering this question and quantify the importance of distinct mechanisms, we mapped a sample of three galactic bulges using data from the integral field
spectrograph WiFeS on the ANU 2.3m telescope in Siding Spring Observatory. Its high resolution gratings (R$\sim$7000) allow us to
present a detailed kinematic and stellar population analysis of their inner
structures with classical and novel techniques. The comparison of those
techniques calls for the necessity of inversion algorithms in order to
understand complex substructures and separate populations. We use line-strength
indices to derive SSP-equivalent ages and metallicities. Additionally, we
use full spectral fitting methods, here the code {\tt STECKMAP}, to extract
their star formation histories. The high quality of
our data allows us to study the 2D distribution of different stellar
populations (i.e. young, intermediate, and old). We can identify their dominant populations based on these
age-discriminated 2D light and mass contribution. In all galactic bulges
studied, at least 50\% of the stellar mass already existed 12 Gyrs
ago, more than currently predicted by simulations. A younger component (age between $\sim$1 to $\sim$8\,Gyrs) is also prominent and its present day distribution seems to be affected much more strongly by morphological structures, especially bars, than the older one. This in-depth analysis of the three bulges supports the
notion of increasing complexity in their evolution, likely to be found in numerous bulge structures if studied at this level of detail, which cannot be achieved by
mergers alone and require a non-negligible contribution of secular evolution. 
\end{abstract}
%=============================================================================


%-----------------------------------------------------------------------------
\begin{table*} 
\centering
\caption{Galaxy sample properties. The columns show the following: (1) NGC number; (2) Hubble type (RC3; \citealt{vaucouleurs1995}); (3) - (6) J2000 coordinates (right ascension, declination), absolute $B$-band magnitude and inclination (taken from HyperLeda); (7) and (8) bulge effective radius and bulge S\'ersic index (taken from \citealt{weinzirl2009}); (9) - (13) H$\beta$ and Mg$b$ line strength indices in \,\AA\, and mag as determined from our data in a central circular aperture of radius of 1.5 arcsec. }
\begin{tabular}{ccccccccccccc}
\hline
Galaxy & RC3 Type & RA & Dec & $M_B$ & incl. & Bulge $r_e$ & Bulge $n$ &
$\sigma_{\rm cen}$ & H$\beta$ & H$\beta$ & Mg$b$ & Mg$b$\\
 ~ & ~ & (h, m, s) & (d, m, s) & (mag) & (deg) & (arcsec) & ~ & (km\,s$^{-1}$)
& (\AA) & (mag) & (\AA) & (mag)\\
(1) & (2) & (3) & (4) & (5) & (6) & (7) & (8) & (9) & (10) & (11) & (12) &
(13) \\
 \hline
NGC\,5701 & (R)SB(rs)0/a & 14 39 11.1 & +05 21 49 & -19.99 & 40.6 & 11.13 &
2.41 &  112 & 3.82 & 0.136 &  1.95  & 0.076\\
NGC\,6753 & (R)SA(r)b & 19 11 23.6 & -57 02 58 & -21.65 & 30.1 & 1.50 & 0.94 &
214 &  4.80  &  0.174 &  1.60  & 0.062 \\
NGC\,7552 & (R')SB(s)ab & 23 16 10.8 & -42 35 05 & -20.52 & 23.6 & 2.70 & 0.64
& 89 & 1.20 & 0.041 & 5.00  & 0.208 \\
\hline
\end{tabular}
\label{tab1}
\end{table*}
%-----------------------------------------------------------------------------

%###############################################################################
\section{Introduction}

Galactic bulges, considered as a deviation from the exponential profile in the
centre of galaxies, are one of the keys to study galaxy formation and evolution
processes, and yet many details of their origin remain unresolved. The different
formation scenarios brought forward during the last decades describe mainly two
bulge types: classical merger-driven bulges and pseudobulges resulting from
secular and/or internal evolution scenarios (e.g. \citealt{korm_rev} and references therein).
Increasingly detailed studies in the last years have however revealed the
presence of rich substructures within those bulges which cannot be fully
attributed to one common evolution scenario. \citet{2005Atha} redefined the
classes based on numerical simulations into: \textit{classical bulges} being
results of mergers or monolithic collapse, \textit{boxy/peanut bulges} formed
via the natural evolution of barred galaxies \citep[see
also][]{1981CombesSanders} and \textit{disc-like bulges} resulting from the
inflow of gas to the centre-most parts triggering star formation. The latter two
bulge types are both results from secular processes within the host
galaxy, so that the division by physical origin remains to be i) classical and
ii) secularly evolved bulges.
 
Pioneering research and recent discoveries show a variety of different bulges
with and without rich substructures hinting to secular evolution - not only
bar-driven. Disentangling these different components can be resolved in
different ways: (i) Photometric observations allow us to study the morphological
features and substructures, e.g. bulge-disc decomposition, through
their light distribution \citep[e.g.][]{2004ApJS..153..411D, 2007Laurikainen} and the derived
broad-band colours can already give us an idea of the present stellar
populations \citep[e.g.][]{bell2000, macarthur2004, 2006GadottideSouza,
munozm2007, Roediger2012}; abundant photometric studies also relate star formation rates and stellar masses to distinct bulge types \citep[e.g.][]{2011FisherDrory} (ii) Spectroscopic observations provide us with the
kinematic properties \citep[e.g.][]{FalconBarroso2006, Ganda2006} and
distribution of stellar populations in these galaxies 
\citep[e.g.][]{Trager2000, Kuntschner2000, macarthur2009, 2011MNRAS.415..709S}.\looseness-2 

While resolving stellar populations would be ideal (instead of integrated light), this is limited to only a few galaxies within the Local Group
\citep[e.g.][]{Tolstoy2009, 2010ApJ...708..560F}. Therefore integrated spectra
and especially colours are usually employed. Thanks to technical developments during the
last decade, the separation of stellar and gas contributions in the spectra
could be achieved \citep[e.g.][]{sarzi2006} and due to better instrumentation,
fainter (sub-)structures could be revealed \citep[e.g.][]{macarthur2009,
Perez2011, 2011MNRAS.415..709S}. Furthermore, major developments in stellar
population analysis techniques coupled with the improved calibration and
extension of spectral stellar libraries (e.g., STELIB,  \citealt{2003LeBorgne}; MILES,  \citealt{2006MNRAS.371..703S};
Indo-US,  \citealt{2004Valdes}; CaT,  \citealt{2001Cenarro1,2001Cenarro2}) have pushed stellar
population analyses forward. 

Early-type galaxies and classical bulges were once assumed to be characterised by single stellar populations whose stars formed long ago on short timescales \citep[e.g.][]{1990Hernquist, Trager2000}. In fact, first bulge studies focused on early-types (to avoid the gas) and compared bulges to elliptical galaxies \citep[e.g.][]{1998Sansom,proctor2002}. However, modern detailed population analyses of bulge systems \citep[e.g.][]{2006MoorthyHoltzmann, 2007Jablonka,morelli2008}, largely based on well-resolved spectroscopy, have revealed a more complex picture of star formation history occurring at both early and later times. Especially the analysis of more complex secularly driven structures which formed over longer periods had to
be revised. For this analysis, inversion algorithms (e.g. {\tt STARLIGHT}: \citealt{cid2005}; {\tt STECKMAP}: \citealt{2006MNRAS.365...74O,  2006MNRAS.365...46O}; {\tt FIT3D}: \citealt{2006NewAR..49..501S}; {\tt ULySS}: \citealt{2009A&A...501.1269K}) were
developed to perform full-spectral fitting of the data comparing it with a
set of synthetic model spectra for a range of ages and metallicities. 

Only very few studies have investigated galactic bulges in this great detail up
to date and only few have used integral-field spectroscopy
\citep[e.g.][]{ganda2007, 2012Yoachim, 2014SB}. \citet{2008AN....329..980O}
demonstrated that a young and cold stellar population could be distinguished
from an old and hot bulge using age - line-of-sight-velocity-distribution
(LOSVD) diagrams. More recent attempts in the literature to achieve similar
goals (using different techniques) are very scarce and usually restricted to
very few, well-known multiple component systems \citep[e.g.][]{2013vdLaan,
2011Coccato, 2013Coccato, 2013Johnston}. Despite great progress, especially with
the advent of large spectroscopic surveys \citep[e.g. ATLAS3D,][]{2011Cappellari}, we are still far from understanding
galactic bulges and their subcomponents, both kinematically and from the stellar
population point of view.

In this paper, we present an in-depth study of three fundamentally different
bulges using the WiFeS integral field spectrograph on the ANU 2.3m telescope in
Siding Spring Observatory. The combination of its large spectral range, high
spectral resolution gratings and sufficiently large field of view allows us to
explore the entire range of tools to derive stellar and gas kinematics, but also
their stellar population content. In Sec.~\ref{sample} we will describe our
target selection and observations, while Sec.~\ref{reduction} provides
details of the data reduction process. The methods employed to perform the
different analyses are described in Sec.\ref{methods}. Section~\ref{kinematics}
presents the results for the stellar and ionised-gas kinematics.
Section~\ref{stepop} introduces the different stellar population results using
the classical indices method and the novel technique via full-spectral fitting.
These results are discussed in Sec.~\ref{discussion}. Finally, 
Sec.~\ref{conclusions} summarizes our main findings. 

%-----------------------------------------------------------------------------
\begin{figure*}
\includegraphics[width=\linewidth]{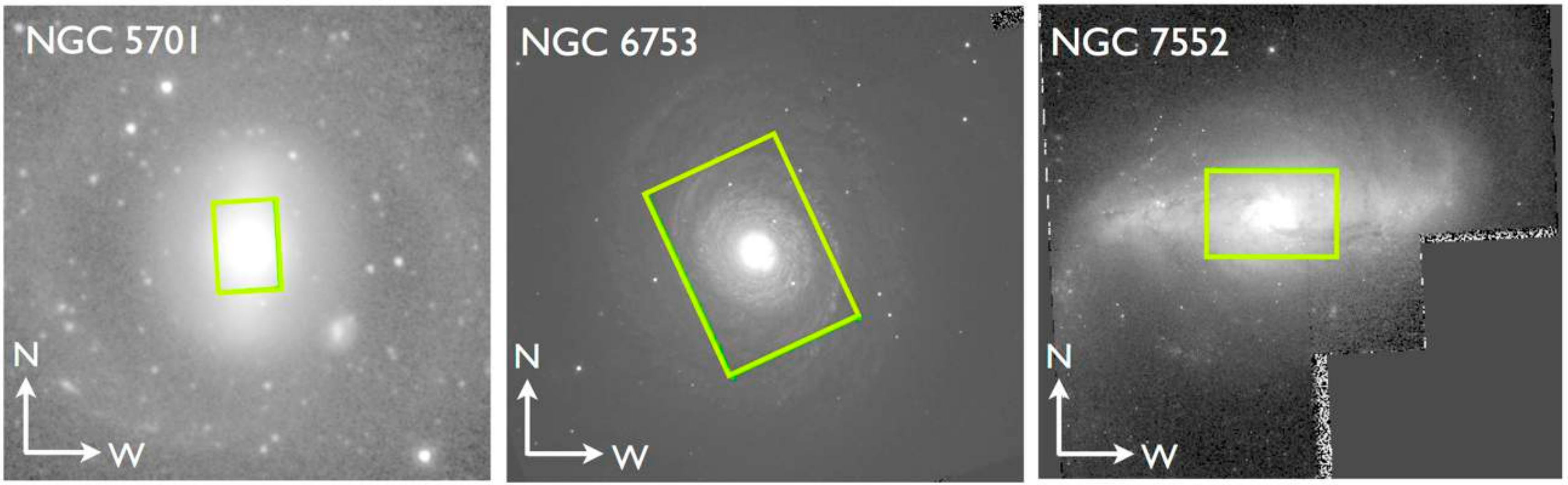}
\includegraphics[width=\linewidth]{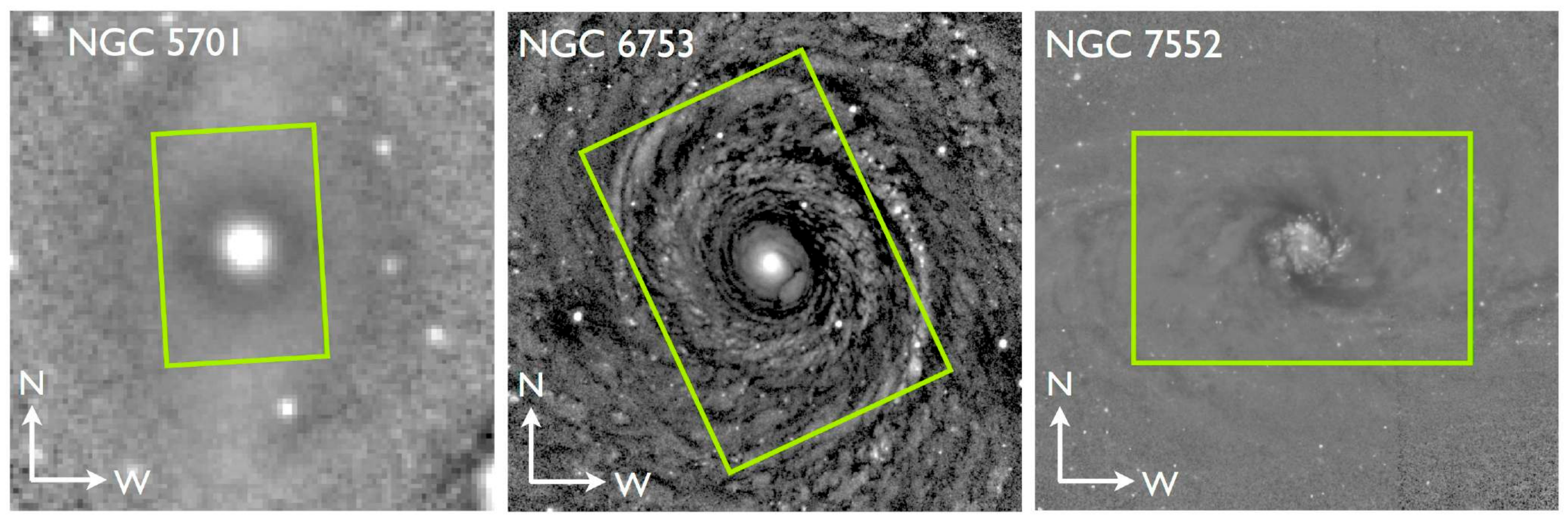}
\caption{(\textit{Top row}) Sample of galaxies observed with the WiFeS
spectrograph. Spitzer 3.6$\mu$m imaging from the S$^4$G survey is shown for
NGC\,5701 (left), while  HST imaging in the F814W filter is presented for
NGC\,6753 (middle) and NGC\,7552 (right). The green rectangle marks the
footprint of the WiFeS FoV (25$\times$38 arcsec). (\textit{Bottom row}) Unsharp mask images for our
sample of galaxies. See \S\ref{sec:mask} for details.}
\label{fig:sample}
\end{figure*}
%-----------------------------------------------------------------------------

%###############################################################################
\section{Sample Observations}
\label{sample}

\subsection{Target selection}

The target selection aimed at providing galaxies with distinct properties and 
level of morphological substructure seen in the photometry in their inner regions in order to quantify
the importance of the different stellar populations present in each system.  
Therefore we chose galaxies with three distinct galactic bulges in barred,
unbarred and ringed galaxies, spanning a very different level of complexity in
stellar populations based on the literature. We selected bright, prominent bulges
to maximize the quality of the data while reducing the required observing time. 
The sample was selected from the Carnegie-Irvine Galaxy Survey \citep{2011Ho}
and the catalogue of inner disks and rings \citep{2002Erwin}, with extensive
ancillary photometric decompositions \citep{weinzirl2009, 2011Li}. Detailed descriptions of each galaxy can be found in the Appendix~\ref{app:galaxies}. Despite the obvious limitations of a sample of only three galaxies, this pilot study allows us to compare these systems and still detect significant similarities, see e.g. \S\ref{sec:results2}, helping us to understand common fundamental formation and evolution mechanisms throughout bulges in disk galaxies. 

We retrieved \textit{Hubble Space Telescope} (HST) Wide Field Planetary Camera 2
archival data, in the F814W filter, for NGC\,6753 and NGC\,7552 from the Hubble
Legacy Archive\footnote{http://hla.stsci.edu, based on observations made with
the NASA/ESA HST, obtained from the European Southern Observatory (ESO)/ST-ECF
Science Archive Facility.}. For NGC\,5701, we used S$^4$G
imaging\footnote{http://irsa.ipac.caltech.edu/data/SPITZER/S4G/}
\citep{2010Sheth} given that no HST data were available (see Fig.~\ref{fig:sample}). The basic characteristics, taken from
HyperLeda\footnote{http://leda.univ-lyon1.fr/}, are listed in Table~\ref{tab1}.
The table also contains bulge characteristics determined by \citet{weinzirl2009}
via two-dimensional surface brightness decomposition as well as central (circular aperture of 1.5 arcsec) velocity
dispersions and line-strength values extracted from our own analysis.\looseness-2

%With the aim to decompose these systems in great detail in this pilot study, we of course acknowledge the limited statistical 

\subsection{Level of substructure from unsharp masking}
\label{sec:mask}
In our galaxies certain substructures are already evident from the photometric
images shown in the upper row of Fig.~\ref{fig:sample}. Nevertheless, we have produced unsharp
masked images, following \citet{Erwin2004} and \citet{Lisker2006a}, to reveal
any small-scale structures or structures with no radial symmetries that may be
present. This method relies on the presence of a smooth and symmetric overall light distribution which can be modeled by the {\tt gauss} task of IRAF\footnote{IRAF is distributed by the National Optical Astronomy Observatory, which is operated by the Association of Universities for Research in Astronomy, Inc., under cooperative agreement with the National Science Foundation.} \citep{1993Tody}. Then the original image can be divided by the smooth model to obtain the unsharp mask. For NGC\,6753 and NGC\,7552, we used a value of
$\sigma_{mask}$\,=\,20 for the Gaussian convolution and $\sigma_{mask}$\,=\,5 for NGC\,5701.
We determined the ellipticity and PA values from our own reconstructed images
(i.e. intensity images extracted from WiFeS datacubes). The results are
presented in the bottom row of Fig.~\ref{fig:sample}. The different substructures stand out very
clearly. NGC\,5701 shows a rather smooth distribution with a strong central
feature. The rectangle indicating the WiFeS FoV exacerbates the proper visualization of its large scale bar, which can be seen rather faint here. NGC\,6753 also exhibits a prominent central component, but with
significant flocculent spiral structure throughout the FoV, but mainly
concentrated in the inner ring. In NGC\,7552, the dust lanes are very evident,
along with the bright circumnuclear ring with star forming regions of different
intensities. 

\subsection{Observations}

The observations were taken in July and September 2013 at the 2.3m telescope at
Siding Spring Observatory (SSO) in Australia. We used the WiFeS IFU which
provides a 38$\times$25\,arcsec$^2$ field-of-view with 1$\times$1\,arcsec per
spatial element. It was commissioned in May 2009 and its detailed description
can be found in \citet{2007Dopita, 2010Dopita}. The instrument's dichroic allows
observations with two gratings simultaneously. Using the RT615 dichroic,  we
chose the two high resolution gratings B7000 and I7000. The B7000 grating
results in a wavelength coverage of 4180 to 5580\,\AA\ with a spectral
resolution ($\sigma$) of 43\,km/s, more details for our data in \S\ref{reduction}. The I7000 grating ranging from 6800 to
8900\,\AA\ supplies the information on the Calcium triplet region. The good
instrumental resolution allowed us to measure the lowest expected velocity
dispersions while the large spectral coverage still ensures a meaningful
full-spectral fitting analysis.  

The central surface brightness for the bulges in our sample is
$\mu$\,$\simeq$\,18\,mag arcsec$^{-2}$ \citep{2011Li}. A minimum S/N per resolution element of S/N$\approx$
20 is required to characterize the kinematics and stellar populations of each
separate stellar components \citep[e.g.][]{2013Johnston}. We aimed at 4 hour
integration times for each galaxy in order to achieve that S/N and still
maintain the maximum spatial sampling provided by the instrument (i.e.
1$\times$1 arcsec). We obtained 4800 seconds (4$\times$20 min exposures) each for NGC\,5701 and NGC\,7552
and 14400 seconds (12$\times$20\,min exposures) for NGC\,6753. Although we lost observation time on the first two targets due to weather conditions, their data are still very useful, just with somewhat coarser binning. The
average seeing was around 1.5\,arcsec, September being slightly better than
July. 

We observed one single pointing per galaxy centred on the bulge dominated
region (see Fig.\,1). Nonetheless, the large FoV allowed us to reach disk
dominated regions. We took calibration frames (bias, flats and arc) before dawn
and after dusk, and sky flats during twilight. The observational strategy was to
``point-and-stare'', i.e. to observe in blocks of object$-$sky$-$object,
calibration frames and calibration stars. This strategy was designed to have sky
and calibrations near each science frame to avoid temporal effects. We decided
to adopt this method rather than the nod-and-shuffle method available for WiFeS
since it maximized the exposure time on the science frames. Instead of taking
the same amount of time on the sky and object frames, we decided to increase the
time spent on the object frames to increase the signal, and at the expense of
slightly larger noise level (i.e. increase by the square root of 2).

%###############################################################################
\section{Data reduction}
\label{reduction}

We reduced and calibrated our data using the new pipeline designed for this
instrument, PyWiFeS\footnote{http://www.mso.anu.edu.au/mjc/wifes.html}. The
pipeline performs a typical reduction on each single WiFeS frame which consists
of 25 slit spectra being 1\,arcsec wide and 38\,arcsec long. The reduction
includes bias subtraction, flatfielding, distortion correction, wavelength
calibration, sky (and additional telluric correction for the red arm)
subtraction, sensitivity curve correction and datacube generation. Details can
be found in \citet{2014Childress}.

For the wavelength calibration of our frames, we had to devise a non-standard
solution. This  was done using neon and argon arc lamp spectra which were
taken close to the science exposures during the entire night. Since this lamp
had not been used before with the B7000 and I7000 high resolution gratings, we
created our own reference files from the arc lamp measurements and calibrated
them with the line values given, relative to air, on the NIST
webpage\footnote{http://www.nist.gov/pml/data/index.cfm}. We ensured the
accuracy of this calibration by reducing arc lamp spectra as well as sky frames
and checking the position of the arc and sky lines. This resulted in an
uncertainty of $\Delta$$\approx$0.1\,\AA. The datacubes were flux calibrated
with the help of flux standard stars observed: HIP71453, EG131 and Feige110, to achieve a relative flux calibration. For
the red-arm spectra, the removal of telluric lines was achieved using
observations of featureless white dwarfs, taken close to the science frames and
at similar air masses as the object. 

The data reduction is run separately for the blue and red arm frames. After
finding the offsets, we use the iraf {\tt imcombine} routine to merge the
individual cubes to a single datacube sampled to a common spatial grid. The
blue spectra span from 4100\,\AA\ to 5500\,\AA\ with a spectral sampling of
0.347\,\AA\ per pixel and a spectral resolution (Full Width Half Maximum, FWHM)
of FWHM$\simeq$0.9\,\AA. The red spectra cover the range from 6808\,\AA\ to
8930\,\AA\ with a spectral sampling of 0.5665\,\AA\ per pixel and a spectral
resolution of FWHM$\simeq$1.5\AA. In both cases the spectral resolution was not
constant along the frame and therefore we convolved the spectra in each case to
the highest measured FWHM values, setting the final spectral resolution to
FWHM$\simeq$1.0\,\AA\ and FWHM$\simeq$1.6\,\AA\ for the blue and red setups
respectively.

%###############################################################################
\section{Methods}
\label{methods}

In our analysis of kinematics and stellar populations, we adopted a Voronoi
binning scheme \citep{2003Cappellari} to reach the desired S/N levels. We chose
to bin our data for NGC\,5701 and NGC\,7552 to S/N$\approx$20 and for NGC\,6753
to S/N$\approx$40. This choice ensures a meaningful analysis while maximizing
the spatial sampling, which is important to resolve substructures present
in our maps. The extension of the field is the WiFeS FoV, however bins of too low signal (less than S/N=3)
have been left out.

\subsection{Stellar kinematics}
\label{stkin}

We extracted the stellar kinematics from the blue and red arm separately. For
simplicity, we present the results from the blue spectra only. Both sets of maps
agree within the uncertainties. We used the pPXF $-$penalized pixel fitting$-$
code developed by \citet{capems_2004} to extract the stellar kinematics. The
routine fits each galaxy spectrum with a combination of template spectra. We
used a subset of PEGASE high resolution model spectra {\tt PEGASE-HR} with
R$\approx$10000 \citep{2004LeBorgne} spanning a wide range of ages and
metallicities in order to minimize the impact of template mismatch. Before the
fitting process, we matched the spectral resolution of those models to that of
our data. pPXF uses a Gauss-Hermite parametrization \citep{gerhard1993,
marel1993} to describe the LOSVD and thus allows the measurement of the velocity
($V$), velocity dispersion ($\sigma$) and higher order Gauss-Hermite moments
(h$_3$ and h$_4$).

\subsection{Ionised gas extraction}
\label{gandalf}

The measurement of the stellar population parameters requires the removal of
the ionised emission present in the spectra. This is particularly important in
the Balmer lines (i.e. H$\gamma$ and H$\beta$) present in our wavelength
range, which are the key features determining the age of the stellar
population.

We use the Gas AND Absorption Line Fitting ({\tt GANDALF}) package by
\citet{sarzi2006} and \citet{FalconBarroso2006} to obtain the ionised-gas
distribution and kinematics. The emission lines are treated as additional
Gaussian templates on top of the stellar continuum and the code iteratively
looks for the best match of their velocities and velocity dispersions. For the
blue-arm spectra, we could measure the following emission lines:
H$\gamma$$\lambda$4341, H$\beta$$\lambda$4861 and the doublets [O{\sc
iii}]$\lambda\lambda$4959, 5007 and [N{\sc i}]$\lambda\lambda$5200, 5202. We
tied spectral lines kinematically as it helps constraining the parameters for {\tt GANDALF} and leaving them free gave matching results. We always chose 
the strongest lines in each case: in
NGC\,5701, we fixed the kinematics of the emission lines to the [O{\sc iii}]
doublet and in NGC\,6753 and NGC\,7552, we fixed the emission
line kinematics to the H$\beta$ line instead. We furthermore imposed known
relative flux relations to constrain the freedom of the lines during the fitting
process, namely $F({\rm H}_{\gamma})=0.469\cdot F({\rm H}_{\beta}$) and
$F([$O{\sc iii}$]_{4959})=0.350\cdot F([$O{\sc iii}$]_{5007})$.

We thus used our results from {\tt GANDALF} to \textit{clean} the spectra
of our galaxies and produce emission-line-free datacubes for our stellar
population analysis.

\subsection{Stellar populations via line-strength indices}
\label{LSmet}
The comparison of absorption line-strength indices measured on observed spectra
with those computed with single stellar populations (SSP) models is the
classical approach to derive stellar population parameters from integrated
spectra \citep[e.g.][]{1973Faber, 1993Davies, 1994Worthey, 1999Vaz, 2005Thomas,
2006SBa, kuntschner2006, macarthur2009}. Lick/IDS indices are the most commonly used to probe
the luminosity-weighted age, metallicity and abundance ratios of specific
elements. 

We started our analysis with this method by relating our index measurements to
the MILES model predictions \citep{2010Vaz}. We obtained the absorption line
strengths in the Line Index System at 8.4 \AA \, (LIS-8.4\AA) \citep{1998Trager, 2010Vaz}.
This approach allows us to avoid the intrinsic uncertainties associated with the
popular Lick/IDS system. In this study, we measured the following indices from the blue grating:
Ca4227, G4300, H${\gamma A}$, H${\gamma F}$, Fe4383, Ca4455, Fe4531, Fe4668,
H$\beta$, H${\beta_o}$ \citep{2009Cervantes}, Fe5015, Fe5270, Mg$b$, Fe5270, Fe5335 and Fe5406. From the red grating, we also determined indices following \citet{2001Cenarro1}: CaT, CaT*, and PaT with Ca1($\lambda\lambda$8484.0-8515.0), Ca2($\lambda\lambda$8522.0-8562.0), Ca3($\lambda\lambda$8642.0-8682.0), Pa1($\lambda\lambda$8461.0-8474.0), Pa2($\lambda\lambda$8577.0-8619.0) and Pa3($\lambda\lambda$8730.0-8772.0).

While we only made use of some of them to derive our results, we provide a summary of
all maps for each galaxy in both gratings in the appendix. In the particular
case of NGC\,6753 the central spectra are intrinsically more broaden than
8.4\,\AA. A few central spectra exhibit an average broadening of
$\approx$9\,\AA. This results in a shift of $\approx$0.02\,\AA\,in H$\beta$ and
$\approx$0.05\,\AA\,in Mg$b$. This would correspond to a difference of 1.5\,Gyr from
our measured value and it is well within our uncertainties. Given the small
effect, and for simplicity, we chose to not convolute the data further and use the
models at 8.4\,\AA. 

In Sect.~\ref{stepop} we use the H${\beta_o}$, Mg$b$, Fe5270 and Fe5335 index
maps to determine the stellar population parameters. We specifically combine the Mg$b$, Fe5270 and Fe5335 indices to obtain the [MgFe]' index
\citep[e.g.][]{2003ThomasMB}, as it is almost insensitive to [Mg/Fe] variations. We
then use the {\tt rmodel}\footnote{
http://www.ucm.es/info/Astrof/software/rmodel/rmodel.html} code
\citep{cardiel2003} to compute the mean luminosity-weighted age and
metallicity. In Fig.~\ref{fig:5701-grid}, we show the index-index diagrams for our
three galaxies with the MILES SSP models for different ages and metallicities
overlaid. Throughout this work we assume a Kroupa initial mass function 
\citep[IMF,][]{2001Kroupa}.

The representation of stellar populations of a galaxy by a SSP is an
oversimplification for spiral galaxies. It is for this reason that this method
is mostly used in systems where we can assume that the locally averaged
metallicity and age do not vary very much across the galaxy (e.g. ellipticals).
The errors when representing the local stellar population by an SSP
are then the same everywhere \citep{peletier2007}. While we do not expect a
single stellar population in many regions of our sample of bulges, this
classical approach provides luminosity-weighted population parameters that are
useful to contrast with the literature of similar objects. 

\subsection{Stellar populations via full-spectral fitting techniques}
\label{steckm}

There are several inversion algorithms in the literature
\citep[e.g.][]{cid2005, 2009A&A...501.1269K} whose main goal is the
reconstruction of the stellar content from an observed spectrum. Full-spectral
fitting techniques allow us to maximize the information encoded in a spectrum as
they use the entire wavelength range and they are not limited to some specific
absorption features (e.g. line-strength indices).

{\tt STECKMAP}\footnote{http://astro.u-strasbg.fr/?ocvirk/} (STEllar Content and Kinematics via Maximum A Posteriori likelihood, \citealt{2006MNRAS.365...74O, 2006MNRAS.365...46O}) is a
full-spectral fitting code that uses a Bayesian method to simultaneously recover
the stellar kinematics and the stellar properties via a maximum a posteriori
algorithm. It is non-parametric so it provides properties such as the stellar
age distribution (SAD) with minimal constraints on their shape. In addition, the
ill-conditioning of the inversion is taken into account through explicit
regularization. 

%-----------------------------------------------------------------------------
\begin{figure*}
\includegraphics[width=0.9\linewidth]{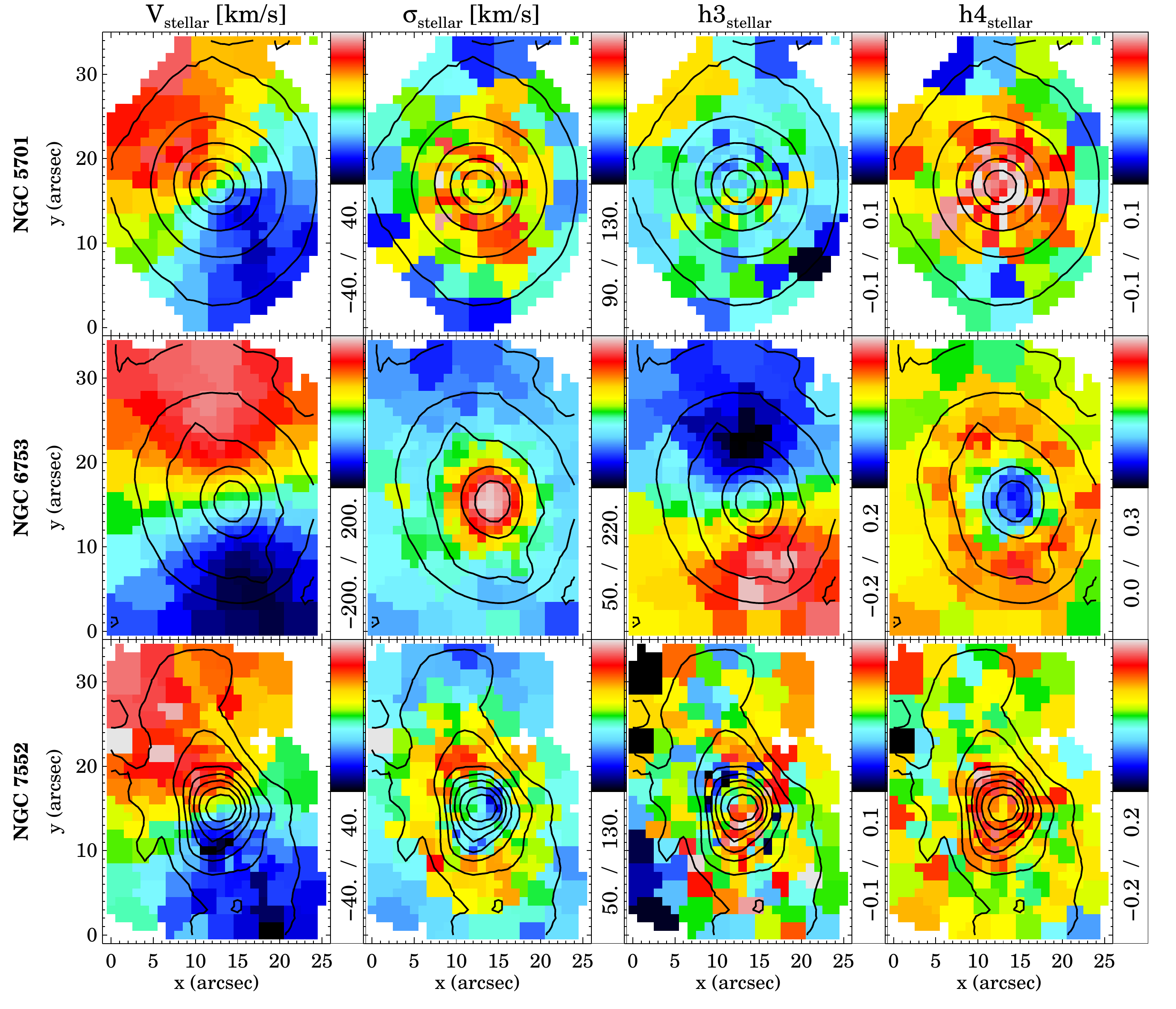}
\caption{Stellar kinematic maps for all three galaxies, from top to bottom:
NGC\,5701, NGC\,6753 and NGC\,7552. For each one, the four panels show stellar
velocity, stellar velocity dispersion, $h_3$ and $h_4$ moments. The colour bars on the side each indicate the range of the
parameter measured. The isophotes shown are derived from the WiFeS cube
reconstructed intensities and are equally spaced in stpdf of 0.5 magnitudes.}
\label{fig:kins}
\end{figure*}
%-----------------------------------------------------------------------------

%-----------------------------------------------------------------------------
\begin{figure*}
\includegraphics[width=0.62\linewidth]{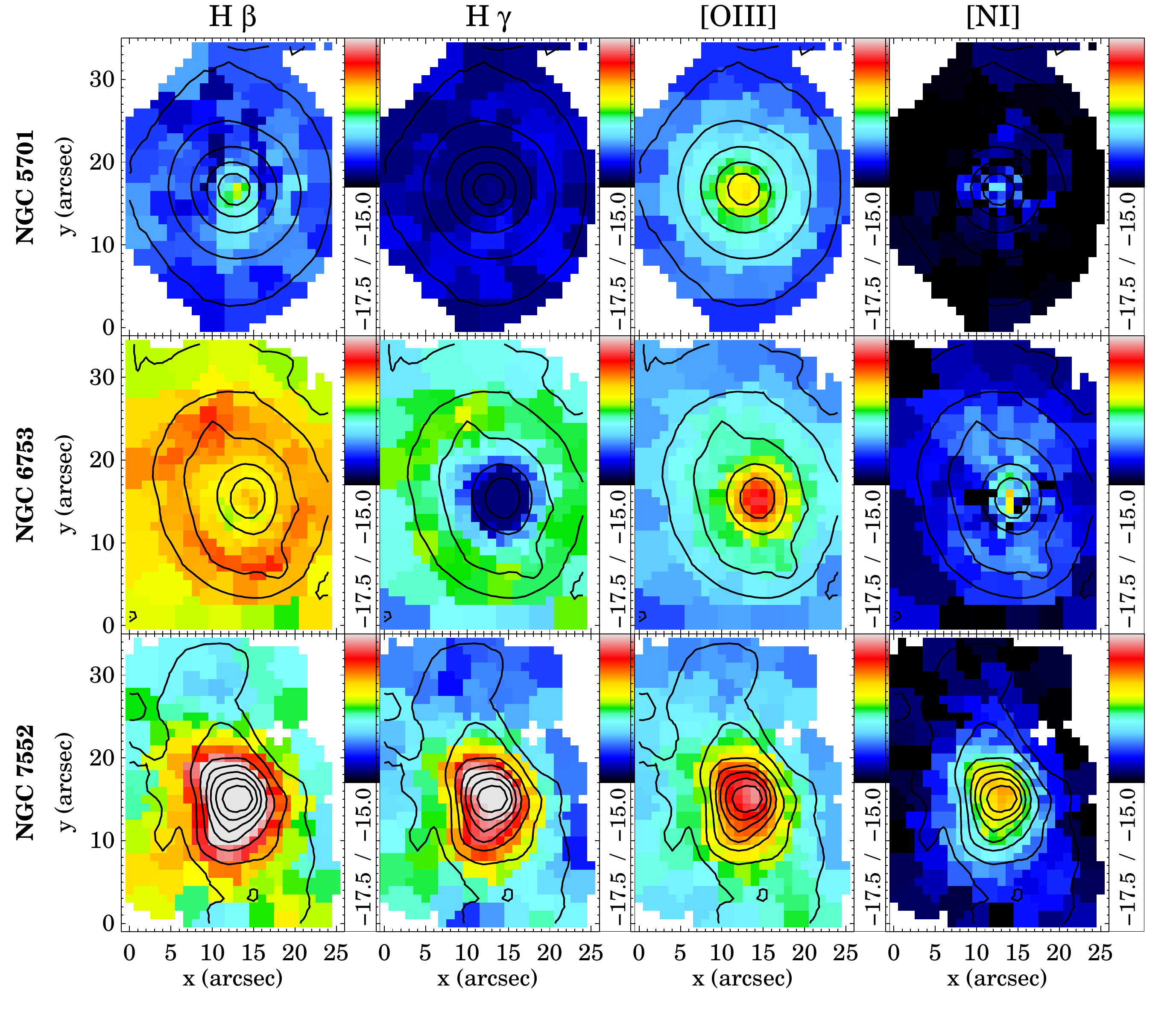}
\includegraphics[width=0.374\linewidth]{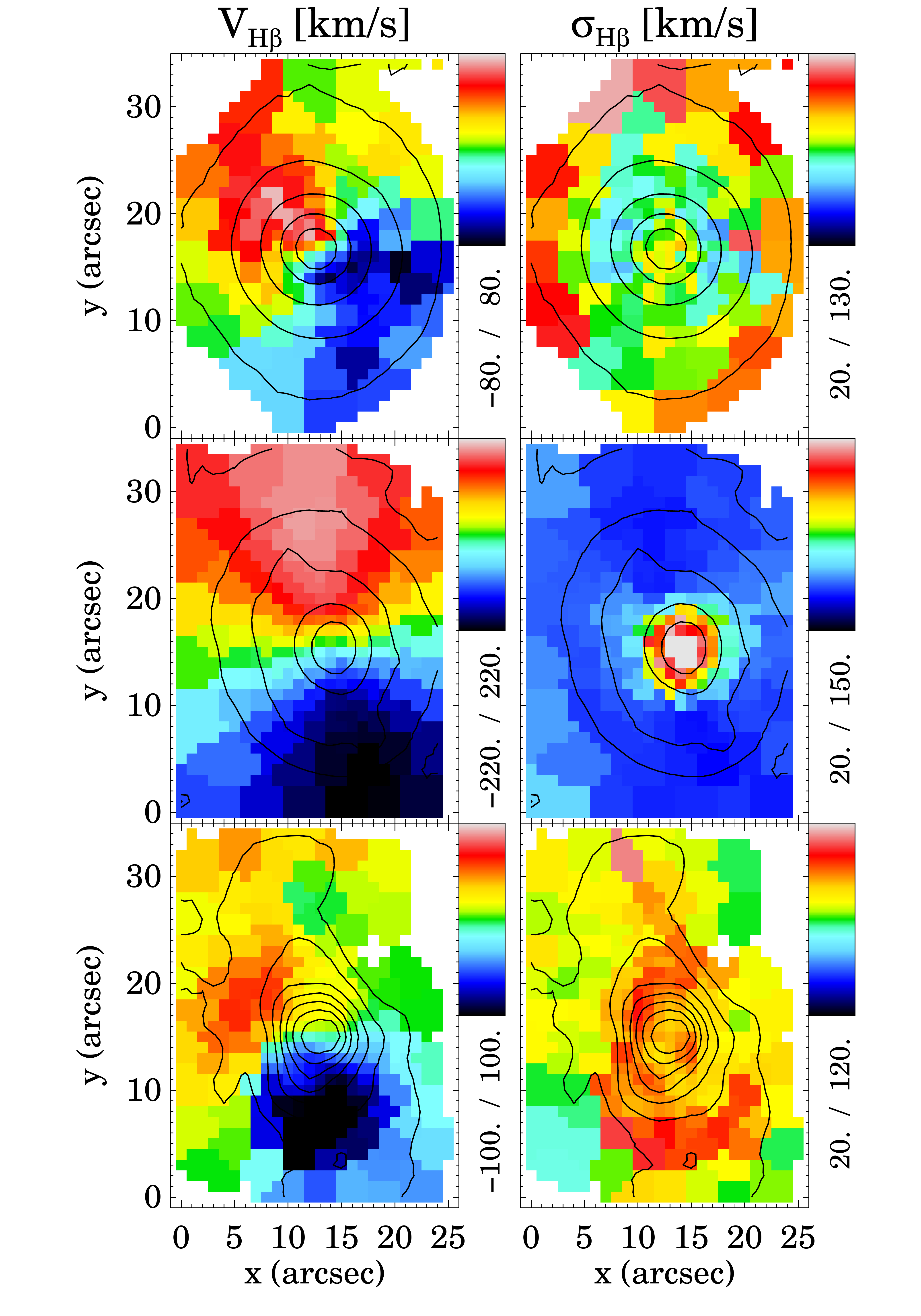}
\caption{Gas fluxes and velocities for all three galaxies in our sample. The
left four columns show the gas intensities, while we present the kinematics on
the right: ionised gas rotation velocity and velocity dispersion. Each row
represents one galaxy, as indicated on the left hand side. NGC\,7552
has a cross of bad pixels on the right (white cross). Fluxes of
the emission lines is given in erg s$^{-1}$ cm$^{-2}$ arcsec$^{-2}$ and in a
logarithmic scale.}
\label{fig:gas}
\end{figure*}
%-----------------------------------------------------------------------------

In practice, the code determines a linear combination of single stellar population models trying  to reproduce the observed spectrum projected onto a temporal sequence of these SSP models. The weights used for the linear combination give the SSP fractions and create the according star formation history associated to the spectrum. Thus, the code does not take any a priori assumption to create the SFHs apart from imposing a smooth solution for the unknown parameters, namely the stellar age distribution, the age-metallicity relation and the line-of-sight velocity distributions or broadening function, which is supposed to avoid un-physical solutions. To achieve this, the code uses certain smoothing parameters whose choice is important, but not sufficiently enough to significantly influence the overall outcome (i.e. main features) of the SFHs, as well as the derived mean values of ages and metallicities. This has been tested in many former works, \citep[e.g.][]{2006MNRAS.365...74O,2006MNRAS.365...46O, 2008AN....329..980O, 2008Kolevab,  2011MNRAS.415..709S, 2011Koleva, 2014SB}. 

For this work, we use the emission-cleaned spectra coming from the {\tt GANDALF} analysis
following the same Voronoi scheme outlined above. We shift the spectra
to rest frame according to the stellar velocity (see Sec.~\ref{stkin})
and broadened them to 8.4\,\AA. We fix the stellar kinematics and fit
exclusively for the stellar content in order to avoid the
metallicity-velocity dispersion degeneracy reported by \citet{2011MNRAS.415..709S}. 

As in the previous section, we use the MILES models as the reference templates with the following range of ages and metallicities: 63 Myrs to 17.8 Gyrs and  -2.32 < [Z/H] < +0.2 respectively. We also keep using the Kroupa Universal IMF. The chosen age range can obviously lead to outputs of ages older than the age of the Universe, but in line with globular cluster ages. Several former studies have investigated this zero point problem \citep[e.g.][]{2001Vaz, 2002Schiavon, 2010Vaz, 2011Maraston} and in order to not artifially biase our outcome, we use the entire range of models available, which is also usually done in SP studies. 

Once we obtain the star formation history of a
given spectrum, we compute the luminosity- (L) and mass-(M) weighted age and metallicity
(both represented by $q$) as follows: 

%-----------------------------------------------------------------------------
\begin{equation}
  \langle q \rangle_{M} = {\sum_i mass(i)q_i}/{\sum_i mass(i)},
\end{equation}

\begin{equation}
  \langle q \rangle_{L} = {\sum_i flux(i)q_i}/{\sum_i flux(i)}.
\end{equation}
%-----------------------------------------------------------------------------

In order to obtain the value of metallicity with respect to solar metallicity $Z_{\odot}$=0.02
we use: 

%-----------------------------------------------------------------------------
\begin{equation}
  [M/H]_{L} = -2.5 \log_{10}(Z_{L}/Z_\odot)
\end{equation}

\begin{equation}
  [M/H]_{M} = -2.5 \log_{10}(Z_{M}/Z_\odot)
\end{equation}
%-----------------------------------------------------------------------------

The comparison of the stellar properties from the line-strength indices and the
full-spectral fitting will allow us to better understand the limitations of the
classical method, i.e. assess the two-fold bias of the indices results proposed
by \citet{serra2007}.

%###############################################################################
\section{Kinematic properties}
\label{kinematics}

Figures~\ref{fig:kins} and~\ref{fig:gas} present maps of the absorption and
emission-line distribution and kinematics of the three galaxies in our
sample. Overlaid in all maps, we show the isophotes of the total intensity
reconstructed from the WiFeS spectra (in mag/arcsec$^2$ with an arbitrary
zero point) equally spaced in intervals of 0.5 magnitudes (detailed maps are collected in Appendix~\ref{app:maps}). 
Here, we concentrate on an overview of the general kinematic trends and results observed.

\subsection{Stellar kinematics}

NGC\,5701 shows a misalignment between the galaxy's main photometric axis and
the bar. The isophotes also indicate that the bar angle is clearly misaligned
with respect to the major kinematic axis. The maximum absolute rotation values
reach up to 40\,km\,s$^{-1}$ within our FoV. The velocity dispersion is higher
in the central parts (110\,km\,s$^{-1}$) and drops down to 95\,km\,s$^{-1}$ at
the edges of the FoV. The highest $\sigma$ values are not found exactly in the
centre, but offset by $\sim$5 arcsec, hence presenting a central $\sigma$-drop. The h$_3$ map reveals some level of anti-correlation
in the central parts with the velocity map. This finding is consistent with the
elevated h$_4$ values in the centre. We expect a dominant bulge in this region
\citep{weinzirl2009}, but also weak nuclear spiral arms \citet{2002Erwin} which
could explain the non-zero h$_3$ and h$_4$ values obtained. 

In NGC\,6753, the photometric and kinematic axes appear to be aligned. This
galaxy displays an unusually large stellar velocity rotation
($\approx$200\,km\,s$^{-1}$) for the assumed inclination of the galaxy
($i$\,$\sim$\,30$^\circ$). It also shows a very high central velocity dispersion
($\approx$214\,km\,s$^{-1}$) that decreases drastically within the inner
kiloparsec. The dispersion map reveals an extraordinary ``hot'' centre embedded
in a significantly colder component, i.e. the disk, with no distinct kinematic
signature of the inner ring. The h$_3$ values anti-correlate strongly with
the velocity values and so does the h$_4$ moment with the stellar velocity
dispersion. The fact that both h$_3$ and h$_4$ values are high in the inner ring
region supports the distinct kinematic properties of this substructure.

NGC\,7552 displays a similar maximum rotation velocity to NGC\,5701 within our
FoV ($\approx$40\,km\,s$^{-1}$). While also hosting a large-scale bar, the line of nodes is almost perpendicular to the bar's position angle, thus the rotation is along the large-scale bar and less misaligned than in NGC\,5701. A closer look reveals enhancements
of the rotation velocity most likely related to the bar. Hence when taking the
profile, we would see the predicted double-hump rotation curve
\citep{2005Bureau}. A high velocity dispersion ring
is clearly revealed outside the circumnuclear ring region. Towards the edges of the field, these values drop.  In the h$_3$ map, only
a slight anti-correlation with respect to the velocity field can be
distinguished in the area where the circumnuclear ring is present. This
anti-correlation is much more apparent in its velocity dispersion versus h$_4$
moment maps.

The three bulges in our sample display a wide range of kinematic features
clearly associated to different photometric substructures, e.g. double-hump
profiles $\sigma$ predicted by simulations of barred galaxies
\citep{2005Bureau}. Particularly interesting is the behaviour of the
Gauss-Hermite higher order moments h$_3$ and h$_4$, which are markedly
different, for all galaxies, in those regions where we expect to find a mixture
of populations. We will use this information in a follow-up paper (Cacho et al.,
in prep.) to extract the kinematic properties of the different stellar
population components present in the centre of these galaxies. 

\subsection{Ionized-gas distribution and kinematics}

We measured the distribution and kinematics of the following emission lines:
H$\gamma$, H$\beta$, [O{\sc iii}] and [N{\sc i}]. The resulting maps are
presented in Fig.~\ref{fig:gas}. 

In NGC\,5701, [O{\sc iii}] is the most prominent gas component. Its flux peaks
in the centre and decreases outwards until it reaches the barred component. We
barely detect [N{\sc i}] and H$_\gamma$. The H$\beta$, however, shows a weak
peak in the centre, compatible with the presence of nuclear spiral structure in
this galaxy \citep{2002Erwin}. NGC\,6753 shows a ring component in the Balmer
lines, clearly visible in the H$\beta$ line map. The upper and lower part of
the ring is strongly enhanced. The [O{\sc iii}] and [N{\sc i}] are mostly
concentrated in the nucleus. NGC\,7552 shows strong central emission in all
probed emission lines, being strongest in H$\beta$ and [O{\sc iii}]. The
H$\beta$ map is in agreement with the H$\alpha$ and radio continuum maps of 
\citet{Pan2013} and \citet{1994Forbesa}, respectively. This comparison
suggests that dust has not affected our measurements significantly. The
very low [O{\sc iii}]/H$\beta$ value in the central regions
($\approx$\,$0.17$) confirms that ionisation is mostly triggered by star
formation \citep[e.g.][]{2001Kewley}.

In the three galaxies, gas rotation velocities are aligned to the
corresponding stellar velocity field. The ionised gas exhibits a higher rotation
velocity than the stars. Conversely, the gas velocity dispersions are lower than
those of the stars. This behaviour is expected given that stars exhibit higher
random motions than the ionised gas. More specifically for each galaxy, we find
that the gas in the centre of NGC\,5701 presents lower velocity dispersion
values than in its outskirts, opposite to the stellar velocity dispersion. The
velocity of the ionised gas in NGC\,6753 is surprisingly close to that of the
stellar kinematics. In NGC\,7552, the gas velocity field shows the same twists
observed in the H{\sc i} and $^{12}$CO (2-1) maps from \citet{Pan2013}. The
velocity dispersion are high around the inner Lindblad resonance (located at a
radius of 1.7 kpc, \citealt{Pan2013}). This region sits just outside the
circumnuclear ring, which has a radius of 0.5\,kpc. The elevated dispersion
values are likely due to shocks induced by the gas arriving at those locations
though the dust lanes along the bar.

%-----------------------------------------------------------------------------
\begin{figure}
\includegraphics[width=1.03\linewidth]{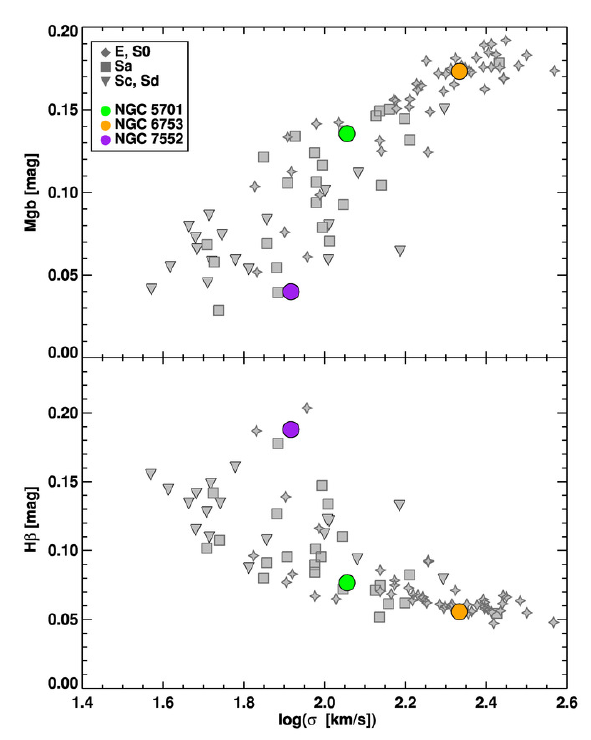}
\caption{Line-index$-\sigma$ relations for our sample of galaxies. Central
aperture measurements of all three bulges (coloured circles) are compared to
values from different samples \citep{ganda2007,peletier2007}. Values for
NGC\,5701, NGC\,6753 and NGC\,7552 are represented by green, yellow and purple
solid circles respectively. Grey symbols show literature values. The upper
panel shows the Mg$b$, expressed in magnitudes, against central velocity
dispersion, while the lower panel shows the central H$\beta$, also expressed in
magnitudes, against velocity dispersion. For our datapoints, we have followed
the conversions from \AA\ to mag described in \citet{kuntschner2006}.}
\label{fig:ind-dis}
\end{figure} 
%-----------------------------------------------------------------------------

 %-----------------------------------------------------------------------------
\begin{figure*}
\includegraphics[width=1.0\linewidth]{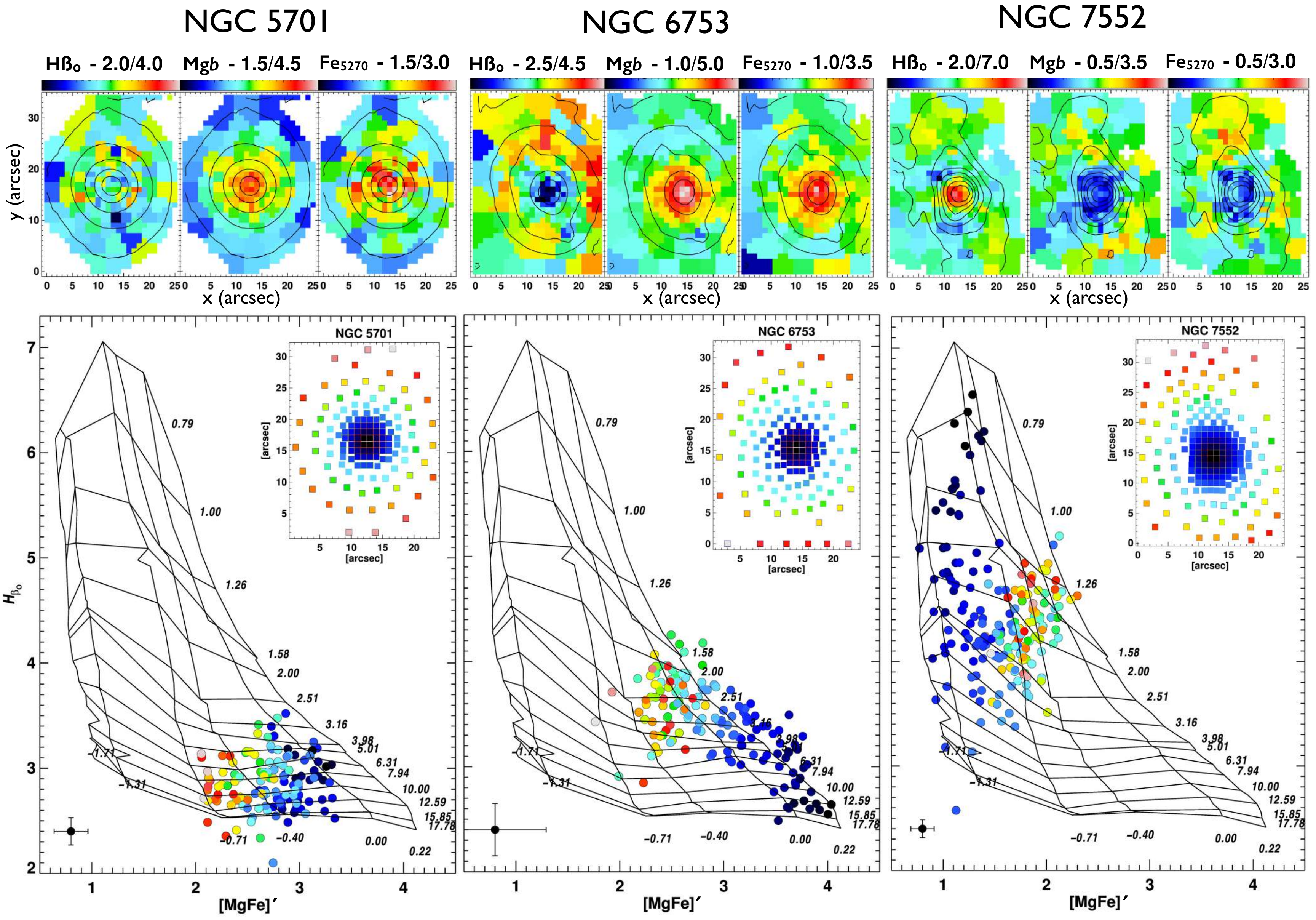}
%\includegraphics[width=1.0\linewidth]{Grids_v10_bw.pdf}
%\vspace{-2cm}
\caption{\textit{Top row:} Maps of absorption line strengths for H${\beta_o}$, Mg$b$ and Fe5270. \textit{Bottom row:} H${\beta_o}$, an age
discriminator, versus the combined index of [MgFe]', indicating metallicities for
the galaxies NGC\,5701, NGC\,6753 and NGC\,7552, from left to right. Overplotted
is a model grid of single stellar populations, roughly indicating the ages and
metallicities (shown on the side of this grid). The points are colourcoded
depending on their distance to the centre of the galaxy and the map in the top
right corners of each panel indicates their position. In the left lower corner we
indicate the typical uncertainty (weighted mean of individual errors) of the points with representative error bars. }
\label{fig:5701-grid}
\end{figure*}
%-----------------------------------------------------------------------------

%###############################################################################
\section{Stellar populations}
\label{stepop}

The rich kinematical substructure found in the previous section may suggest 
a similar variety in the stellar populations of our galaxies. As a first test to classify them we
have compared the central properties with larger samples in the well-known 
line-index$-\sigma$ relation. This relation is well established for
elliptical galaxies \citep[e.g.][]{1981Terlevich}. In recent surveys, e.g.
SAURON survey, this relation was confirmed for early type spirals
\citep[e.g.][]{peletier2007} and extended for late-type galaxies, which showed
larger scatter \citep[e.g.][]{ganda2007}. 

In order to check if our small sample could contain any atypical galaxy which would be unrepresentative of its type, we compared it to other measurements in the literature of similar galaxies.
Figure~\ref{fig:ind-dis} shows our measurements of central apertures (1.5\,arcsec, same aperture as in \citealt{ganda2007}) in comparison with those in the
literature. NGC\,5701, indicated by a green circle, lies on the edge
between the E, S0s and Sa galaxies on the relation. Despite its large-scale bar,
its central bulge parameters resemble a bulge of any early-type galaxy.
NGC\,6753, indicated by a yellow circle, lies exactly on the cloud of E-S0
galaxies, which may be surprising given the presence of spiral structure in the
inner parts. Its centre is thus similar to classical S0/Sa type bulges.
NGC\,7552, shown by the purple circle, follows the behaviour of late-type
galaxies, likely due to the prominent central starburst. Our sample of bulges contains, at least in their central stellar content, examples of the wide
population of nearby galaxies.

\subsection{Classical index$-$index diagrams}

Figure~\ref{fig:5701-grid} presents the measurements of absorption line
strengths in index-index diagrams. The top row displays the line-strength index
maps for each galaxy: H${\beta_o}$, as an age indicator, and Mg$b$ and Fe5270
as proxies for metallicity. In the second row, we plot H${\beta_o}$ against
the combined index of magnesium and iron, [MgFe]' (using Mg$b$, Fe5270 and Fe5335, see \S~\ref{LSmet}), and overplot a grid of MILES
single stellar population models for Kroupa IMF.  

NGC\,5701 appears to be the oldest galaxy of the three, with a large
scatter in age - from about 3\,Gyr to 15\,Gyr - at almost all radii. The error bar does not account for this observed spread, but the overlapping structures of bulge, bar and disk might lead to this variation.
While this age spread seems to be independent of radius, the metallicity of the
stars is clearly higher towards the centre, even considering the error bar. In NGC\,6753, we find a steep
gradient in age from the very central parts towards the ring (green points) and
then an almost flat behaviour until the edges of the field. The galaxy hosts an
old metal-rich population in the centre, but as soon as we enter the region
dominated by the circumnuclear ring, those measurement points decrease
excessively in age and most importantly fall outside of the model grid.
NGC\,7552 shows the opposite behaviour in the age of the central population. The
central component is very young, as expected in a starburst galaxy. The known
inner ring (of about 5\,arcsec radius) is so small that we cannot
distinguish it from the centre. It is interesting to note two low H${\beta_o}$
regions above and below the centre. These locations correspond to the contact
point of the gas and dust lanes with the inner ring. While the low values can be
the result of dust affecting our measurements, it is also possible, as observed
by e.g. \citet{2008Boeker}, that star formation is suppressed in those
contact points and only enhanced once the gas enters the ring.

This classical approach of measuring stellar population parameters, while in
principle valid for some of the regions in our galaxies, presents a number of
important shortcomings. The most notable is the surprisingly large number of
points in NGC\,6753 that fall outside the grid. As we demonstrate in
Appendix~\ref{milesmodtest}, this is likely due to the complex mixture of
populations present in those regions. This kind of bias makes it
impossible to determine accurately stellar population parameters such as [Mg/Fe]
in many locations of our galaxies. [Mg/Fe] is particularly interesting as it
serves as a chemical clock to establish the speed of a star formation event
(i.e. being higher for quick star formation episodes). In the remaining of this
paper we will only determine and use the information provided by this ratio in
areas which are mostly dominated by a single stellar population (see Sec.~\ref{discussion} for its determination).

%-----------------------------------------------------------------------------
\begin{figure*}
\includegraphics[width=0.96\linewidth]{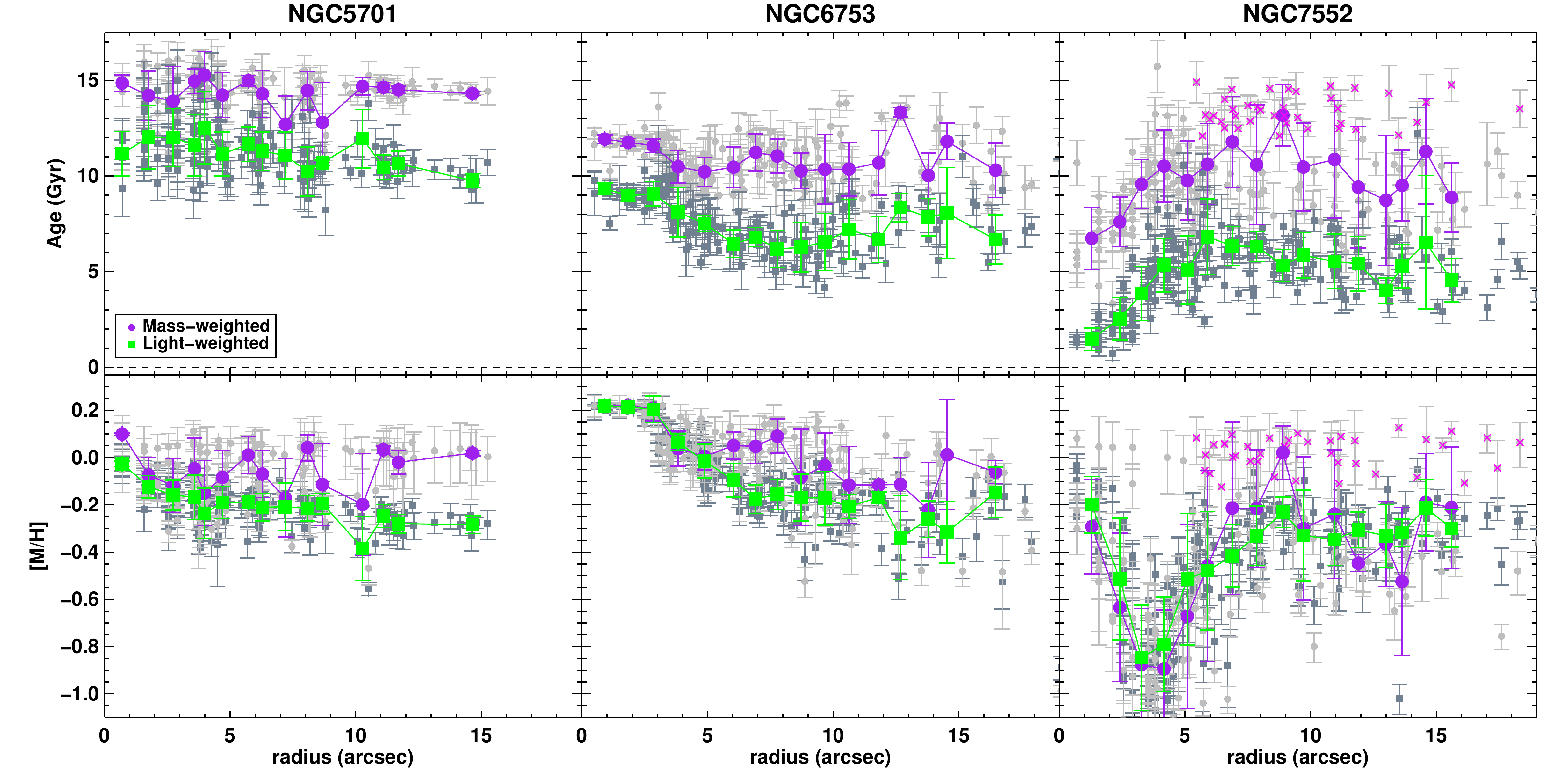}
\caption{Stellar age and metallicity profiles for all three galaxies obtained
with {\tt STECKMAP}. The first row shows the age profiles. In the second row the
metallicity trends. The luminosity-weighted values (green, with individual points in dark grey) and mass-weighted
values (purple, with individual points in light grey) are overplotted. Errors were computed through Monte Carlo simulations. A cloud of interesting individual points for M-weighted results in NGC\,7552 is marked by pink crosses (see text for details).}
\label{fig:SM-TRL}
\end{figure*}
%-----------------------------------------------------------------------------

\subsection{Radial stellar populations from full-spectral fitting}

We obtained ages and metallicities with $rmodel$ from the indices as well as
luminosity- and mass-weighted values from the full-spectral fitting with {\tt
STECKMAP}. For simplicity, given that the index results are similar to the
light-weighted results, in this section we only present the radial profiles of
the mean stellar age and metallicity measured with {\tt STECKMAP}. In
Fig.~\ref{fig:SM-TRL}, we plot the mean stellar age and metallicity trends (both
luminosity- and mass-weighted) together with the cloud of individual values in
our maps. The relations are computed as the median of the individual values
found in every Voronoi bin over 1\,arcsec annuli. The $[M/H]$ is determined from
the metallicity values that {\tt STECKMAP} gives using a solar metallicity
reference of $Z_{\odot}$=0.02. 

We calculated the uncertainties in the parameters through a series of 25 Monte Carlo (MC) simulations. We tested the difference of 25 versus 250 MC simulations and found the resulting errors to be the same within 1-2\% difference. In detail, the MC procedure employed is as follows: once the best fit for the best age and metallicity values is obtained, we create 25 mock spectra by adding noise to this best fit matching the S/N of the observed spectrum. Then we run STECKMAP on those mock spectra using flat first guesses for the Stellar Age Distribution (SAD). The age (metallicity) error is the standard deviation of the ages (metallicities) of the mock fits.
NGC\,5701 displays a rather flat age profile, both luminosity- and
mass-weighted. The metallicity profile goes from solar to sub-solar values (for
the luminosity-weighted values) from the centre to the outer parts. As expected,
the luminosity-weighted trends found here are quite similar to the ones obtained
via line-strength indices (compare with the index values, Fig.~\ref{fig:5701-grid}).
The comparison of those trends with the mass-weighted results suggests an
uniform stellar distribution in the field of view of our data. Particularly
interesting is the difference between the luminosity- and mass-weighted results in
the very centre: the mass-weighted age being high, whereas the
luminosity-weighted age shows a slight drop. The decrease in age could be
due to the nuclear spiral structure present in that region.

NGC\,6753 shows a much richer behaviour, suggesting a more complex stellar
content. This galaxy contains a circumnuclear ring of young stars between 
5\,arcsec and 10\,arcsec. The centre of this galaxy is quite old
(both in the L- and M-weighted sense). The M-weighted age profile is flatter
than the L-weighted. The metallicity profile saturates at the centre (i.e. 0.22
is the most metal-rich population in the MILES models) and shows a steep
negative gradient until it reaches the ring, where the profile flattens
($[M/H]$\,$\sim$\,$-0.2$ for the L-weighted and solar values for the M-weighted
profiles). The apparent broadening of the lines due to the very high central velocity dispersion did not exacerbate the STECKMAP results, since the kinematics are given by ppxf and the resulting fits are very reasonable. 

%-----------------------------------------------------------------------------
\begin{table*}
\centering
\caption{Three example test results as an excerpt of our test series. Here we are using i) a constant SFR (Test example 1) ii) exponential declining SFRs (Test examples 2 and 3), producing ranges of young-intermediate-old fractions compatible with what we might expect for real galaxies. The input is given as a mass fraction and can directly be compared with the M-weighted value which we recover with {\tt STECKMAP} in the same way as for our data. The SFR input is to be compared with the determined SFR and can be related to the L-weighted value.} 
\begin{tabular}{l||||rrcl|||||||||||||||||||rrcl}
\hline
\textbf{Test} & ~  & \textbf{Young} & ~ & ~ & ~ & \textbf{Intermediate} & ~ & ~    \\
 ~ & Input (Mass) & Input (SFR)  & M-weight & SFR & Input (Mass) & Input (SFR)  & M-weight & SFR \\  
 \hline
\textbf{1} & 0.05 & 0.27 & 0.04$\pm$0.01 & 0.29$\pm$0.04  & 0.44  & 0.53 & 0.40$\pm$0.13 & 0.47$\pm$0.10 \\
\textbf{2} & 0.00 & 0.00 & 0.00$\pm$0.00 & 0.00$\pm$0.01  & 0.01  & 0.03 & 0.07$\pm$0.07 &  0.14$\pm$0.12 \\
\textbf{3} & 0.03 & 0.38 & 0.03$\pm$0.02 & 0.32$\pm$0.06  & 0.26  & 0.32 & 0.27$\pm$0.15 &  0.33$\pm$0.13 \\
\hline
\hline
\textbf{Test} & ~ & \textbf{Old} & ~& ~ & ~ & \textbf{Extra Old} & ~& ~ \\
 ~ & Input (Mass) & Input (SFR)  & M-weight & SFR &  Input (Mass) & Input (SFR)  & M-weight & SFR \\ 
 \hline
 \textbf{1} & 0.13 & 0.07 & 0.12$\pm$0.02 & 0.07$\pm$0.01  & 0.38  & 0.13 & 0.44$\pm$0.14& 0.17$\pm$0.08 \\
 \textbf{2} & 0.03 & 0.04 & 0.06$\pm$0.04 & 0.09$\pm$0.04  & 0.96  & 0.93 & 0.87$\pm$0.10& 0.77$\pm$0.15 \\
 \textbf{3} & 0.13 & 0.06 & 0.15$\pm$0.03 & 0.10$\pm$0.02  & 0.58  & 0.22 & 0.55$\pm$0.17& 0.25$\pm$0.12 \\
\end{tabular}
\label{tab:LMtests}
\end{table*}
%-----------------------------------------------------------------------------

Young stars ($\sim$1\,Gyr) are found at the centre of NGC\,7552, followed by a
sudden increase in age until $\sim$6\,arcsec where the profile appears to
flatten. The M-weighted age profile behaves similarly, but does not show such
young populations in the nuclear region. The L-weighted (M-weighted) metallicity
in the centre of this galaxy is below solar followed by a sudden drop to a value
of $[M/H]\approx-1.0$, followed by another gradual increase towards
$[M/H]\approx-0.2$ values up until $\sim$8\,arcsec, from where it stays
constant. The young and metal-rich stars that we find in the centre of this
galaxy are consistent with the central starburst reported for this
galaxy  \citep{1994Forbesb, Schinnerer1997, Pan2013}. Outside the inner\,5 arcsec, the galaxy displays values similar
to those of NGC\,5701 and NGC\,6753 at the same radii. Our mass-weighted profile shows an
interesting feature both in age and metallicity: a separate cloud of points
older than 12\,Gyr and around [M/H]=0.0 dex values at radii larger than
5\,arcsec (marked with pink crosses in the figure). We investigated the location of these bins in our maps and they belong to regions in the ring where the velocity dispersion is large (see
Fig.~\ref{fig:kins}) and hence might represent a distinct population with clearly different kinematics.

Two of our three galaxies host known circumnuclear star-forming rings and
therefore the presence of the young stars detected from our data is not
unexpected. The current analysis so far has focused in average luminosity- or
mass-weighted quantities and therefore do not necessarily reveal, specially in
the ring-dominated regions, the presence of any underlying old stellar
population. Earlier studies of the stellar populations in star-forming rings
\citep[e.g.][]{Allard2006, 2013A&A...551A..81V} have found a non-negligible amount of
old stars in the ring regions (the mass fraction of young stars in the ring is only 30-40\%). In the next section we
will take advantage of the possibility {\tt STECKMAP} gives us to decompose the
stellar populations of our bulges into their main constituents to establish the
amount of old, intermediate and young populations present in them. We will use
that information, together with evolutionary models, to set constraints on the
level of secular versus merger driven processes taking place in our galaxies.

%-----------------------------------------------------------------------------
\begin{figure*}
\includegraphics[width=0.7\linewidth]{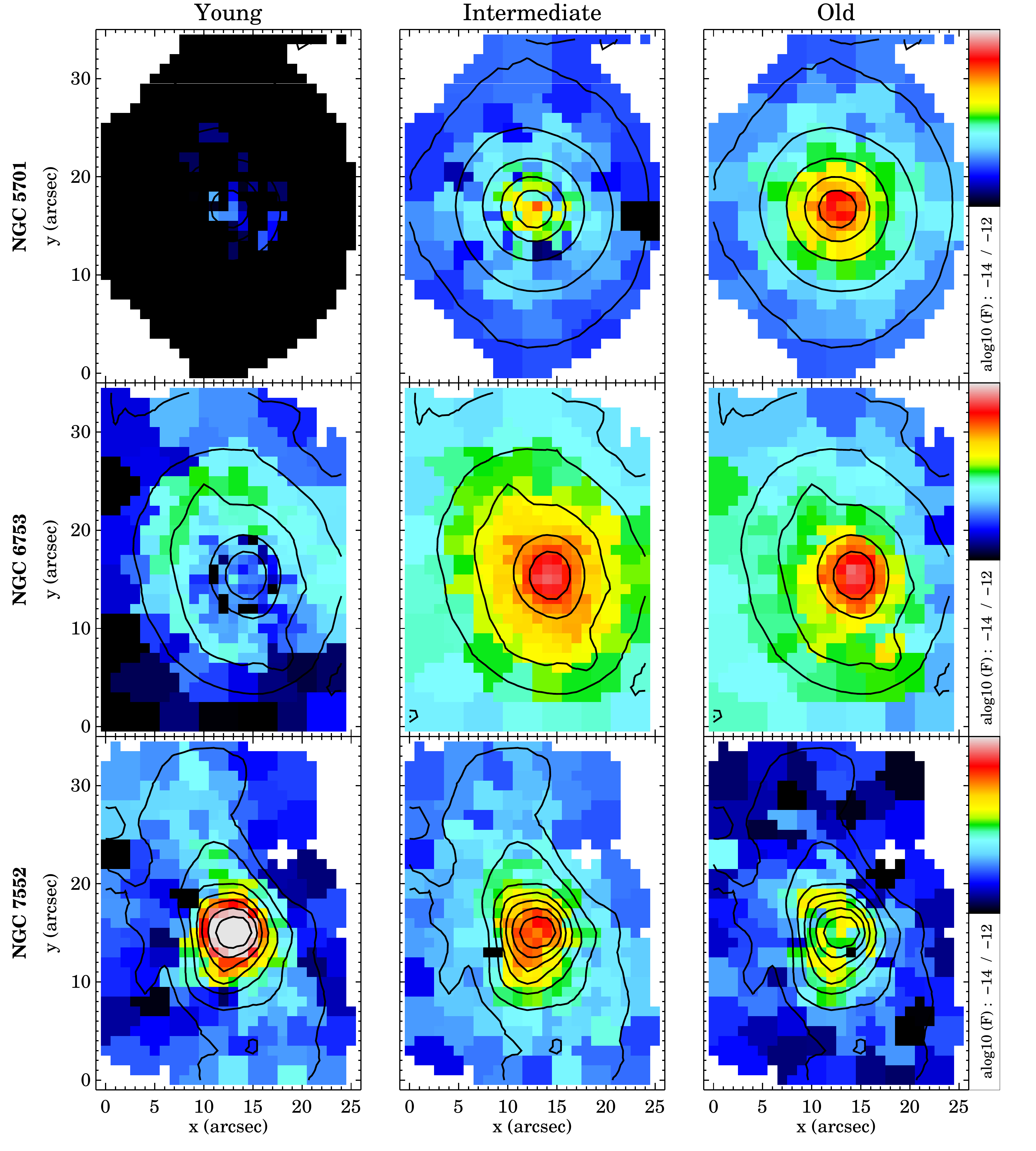}
\includegraphics[width=0.285\linewidth]{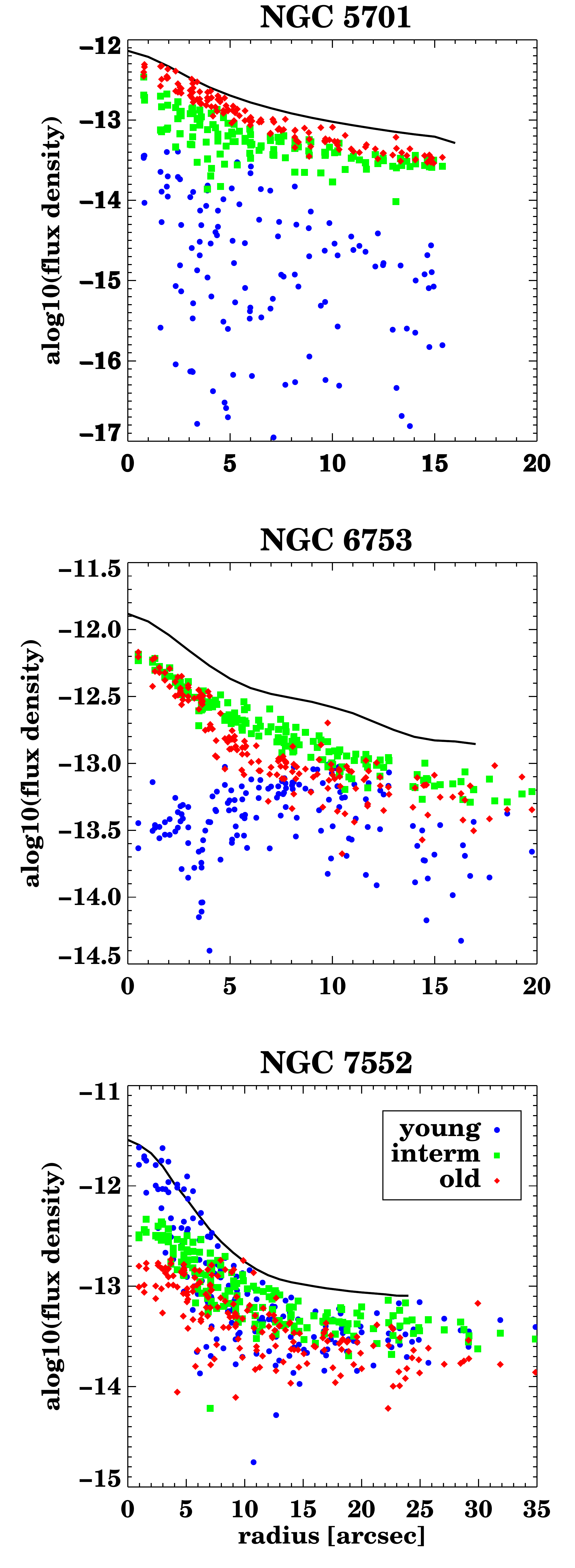}
\caption{Surface brightness maps showing the relative contribution of young,
intermediate and old stars in each spatial element throughout the galaxies and
stellar luminosity profiles revealing the radial contribution of each
component (young (blue circles), intermediate (green squares) and old (red
diamonds). The black line indicates an ellipse fit performed with IRAF on the
intensity image obtained with WiFeS. Fluxes are given in erg s$^{-1}$ cm$^{-2}$
arcsec$^{-2}$ and in a logarithmic scale.}
\label{SM-Lum-profiles}
\end{figure*}
%-----------------------------------------------------------------------------

%###############################################################################
\section{Dissecting the stellar content and its implications}
\label{discussion}

The observational data clearly suggest the presence of different stellar
populations and demonstrate their complexity likely present in general in galactic bulges when studied in great detail. With the aid of {\tt STECKMAP}, we separated the
different population components, both L- and M-weighted, in three age bins: young  ($\lesssim$\,1.5\,Gyr, formation redshift $z$ $\lesssim$ 0.1),
intermediate (1.5\,Gyr\,$\lesssim$\,intermediate\,$\lesssim$\,10\,Gyr, 0.1 $\lesssim$ $z$ $\lesssim$ 2) and old ($\gtrsim$\,10\,Gyr, $z$\,$\gtrsim$\,2). A visualization on how this is achieved can be found in the appendix, section~\ref{app:explots}. For the conversion between ages and formation redshift, we are using a standard $\Lambda$CDM (cold dark matter) cosmology with a Hubble constant of  H$_0 = 68.14$ km/s/Mpc and a value of the matter density parameter of $\Omega_m = 0.3036$.

Our aim is to reveal their spatial distribution within the galactic bulges and
understand how different star formation epochs (associated to the distinct age
cuts) influenced the evolutionary histories of these galaxies. The review on cosmic SFH \citep{2014MadauD} summarizes distinct scenarios according to different epochs, which we will discuss more in \S\ref{results3}. We are conscious about the oldest age of SSP models exceeding the age of the universe. This has been detected in former studies \citep[e.g.][]{2001Vaz} and is mainly due to degeneracies (age, metallicity, IMF, etc.) in old systems and using these models does not change the cosmology. 

We acknowledge the increasing difficulty of separating intermediate and old stellar populations,  but STECKMAP has been extensively tested in different works \citep{2006MNRAS.365...74O,2006MNRAS.365...46O, 2008AN....329..980O, 2008Kolevab,  2011MNRAS.415..709S, 2011Koleva}. The strategy and set of parameters used in this paper while running STECKMAP are the result of a series of tests following different schemes by different groups and by our own  \citep[e.g.][]{2014SB}. Furthermore we point out that the age cuts are an orientation and should be taken as an age range rather than a clear cut. 

Additionally, we performed our own test series using combinations of model spectra according to our age cuts and recovered their L- and M-weighted age fractions within our proposed cuts with {\tt STECKMAP}. Table~\ref{tab:LMtests} shows the quantitative results for three tests, the first using a constant SFR and the second and third using exponential SFRs. We also tested a combination of bursts using inputs similar to the mass fractions we obtained for the galaxies and also recovered those inputs. In all cases we see that a negligible mass of young population still causes an appreciable fraction in light, while the intermediate and old component are more dominant in mass. In fact if the mass of this young component is high (more than 10\%), it will contribute a lot to the light and the old fraction can be underestimated (we observed this in other tests). Since the young mass fraction in none of our galaxies exceeds 4\% (and this only in the centre of NGC~7552, in all the rest it is well below 1\%), we assume that our measurements resemble more the test cases we show (and similar) and are therefore reliable. We point out that in the tests shown, the old population does not correspond to the entire old population (formation redshift z>2) but only the fraction until the extra old population (z>4), thus 2<z<4.

\subsection{Surface brightness profiles for each sub-population}

From the different weights of each stellar populations given by {\tt STECKMAP},
we can derive their contribution to the overall light of the galaxies. We used 
the reconstructed surface brightness distribution and multiply with the
luminosity-weighted maps for each sub-population. The result of this exercise
is shown in the form of maps and radial profiles in
Fig.~\ref{SM-Lum-profiles}. 

NGC\,5701 is dominated by the light of an old stellar population, while in its
outer parts, an intermediate population gains in importance. In the maps of
Fig.~\ref{SM-Lum-profiles}, we clearly detect the contribution of the central
nuclear spirals in the young component (12\% of the light), while
the overall luminosity profile is clearly dominated by the population formed
long ago. Young populations often outshine old components, but here even though
a young component exists, the old remains dominant. It is already evident that
the young fraction must be negligible in mass (as we will show later).

The light of NGC\,6753 originates mainly from an intermediate-age population
with a significant contribution in light from an old population, mostly in the
centre. As seen in the reconstructed maps, the contribution in light of the
young population in the ring region is considerable (i.e. as much as the old
component). At large radii no clear morphological feature can be associated to any of the populations.

NGC\,7552 shows the young population as the major contributor to the surface
brightness distribution, particularly in the central region but also along the
bar. The contribution of the intermediate-age population is still significant,
being larger than the old component at all radii. Surprisingly a
non-negligible amount of old material (almost $\sim$ 30\%) is organised in a
ring-like structure. This is unlike the ring in NGC\,6753 where most of the
material in the ring comes from an intermediate-age population. The Hubble image
of NGC\,7552 shows that the ring is not a closed structure, being brighter
North from the nucleus. This feature is also observed in our analysis as the
ring exhibits an age gradient, being younger on those regions. The young
populations distributed along the bar suggests that gas must have been 
funnelled towards the central regions sustaining star formation over a long
period.

%-----------------------------------------------------------------------------
\begin{figure*}
\includegraphics[width=1.0\linewidth]{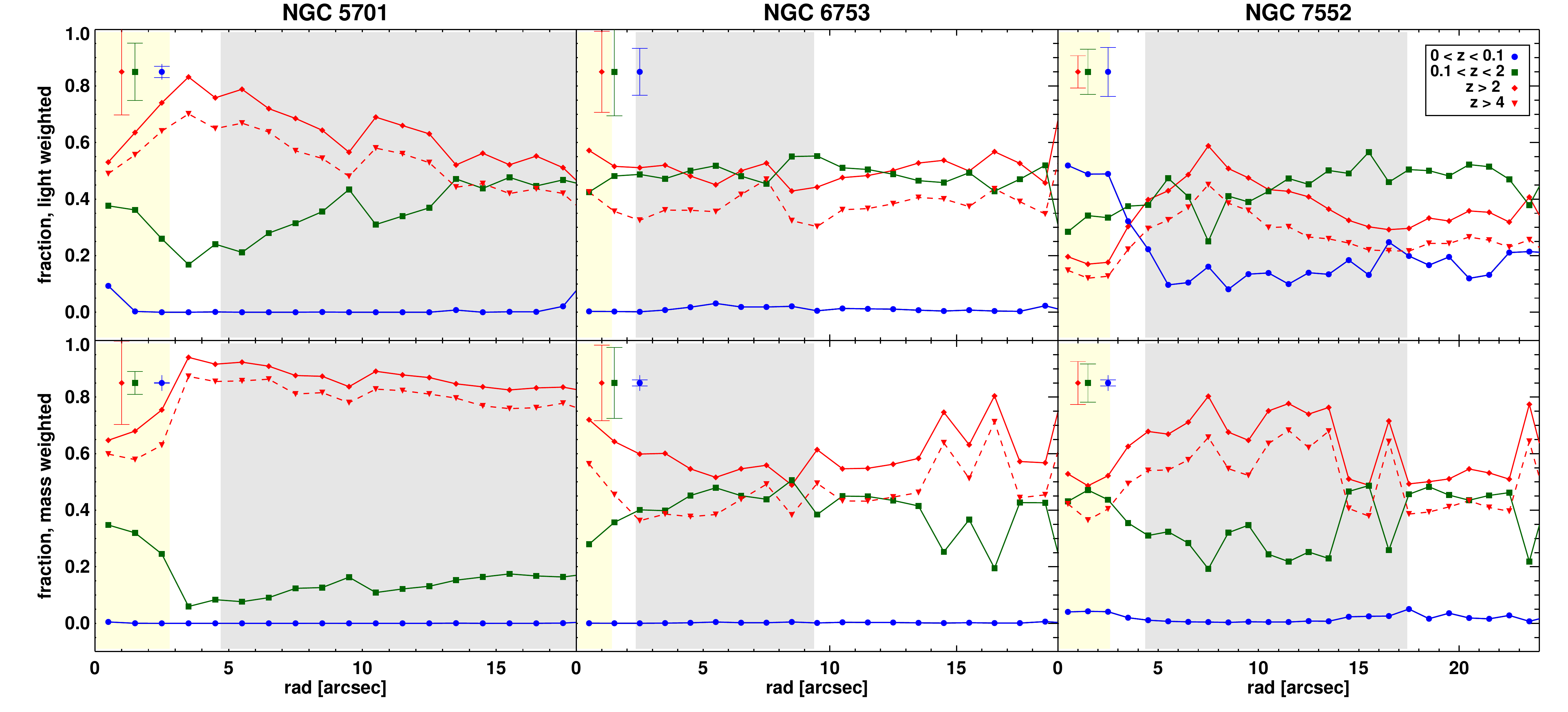}
\caption{Fraction of young (blue dots), intermediate (green squares) and old stellar population (red rhombus, even older: red triangle and dashed line) as a function of radius. In the top left corner we indicate the corresponding uncertainties. \textit{Top row:} L-weighted quantities; \textit{Bottom row:} M-weighted results. Shaded regions indicate the central (< 0.3 kpc) and inner parts (0.5 kpc < r < 2 kpc) where we determine average contributions of each population (see Tab.~\ref{tab2}).}
\label{SFHs}
\end{figure*}
%-----------------------------------------------------------------------------

\subsection{Stellar age distributions and mass content} 
\label{sec:results2}

Figure ~\ref{SFHs} displays radial profiles of the distinct age fractions, both L- and M-weighted. We separated once more into young, intermediate and old populations, additionally indicating an even older epoch by a dashed line. The motivation of this separation remains the same: the attempt to distinguish between different processes which according to theory happened at distinct epochs of the universe. The additional older age cut was motivated by the high mass fraction found in the former old age cut. We thus tried to constrain the formation redshift even further to compare with mass fractions proposed by cosmological models. 

Here we binned the spectra in ellipses in order to raise the S/N in each of them to obtain radial SFHs and from those the distinct age fractions (a visualization of the two binning schemes can be found in the appendix, section~\ref{app:explots}). The ellipticity was determined from reconstructed images directly from our WiFeS data cubes, using the IDL routine {\it find\_galaxy.pro} written by Michele Cappellari and available as part of the mge\_fit\_sectors package\footnote{http://www-astro.physics.ox.ac.uk/$\sim$mxc/idl/}. Table~\ref{tab2} summarizes the fractions of each of these populations for the central and entire inner parts. We also point out that the L-weighted quantities here (Fig.~\ref{SFHs}) are not only obtained with a distinct binning scheme but are different measures than the computed light profiles associated to the different populations, shown in Fig.~\ref{SM-Lum-profiles}. Please see the visualization shown in Fig.~\ref{Ex1} for details.

The upper panel in row 1 of Fig.~\ref{SFHs} already reveals at first sight that the bulge of NGC\,5701 is dominated by old stars throughout. These L-weighted quantities show the old centre and at around 13 arcsec, the intermediate population starts to reach the same luminosity as the old component.  A slightly younger age can only be distinguished in the central 2-3 arcsec ($\sim$10\% in the central bin, significant enough considering the small error bar for the young population). In mass, this population of $\le$1.4 Gyrs is negligible . This is likely the light contribution of the nuclear spiral structure. The M-weighted results show that $\sim$70-85\% of the stellar mass existed in fact already at $z$$\sim$2. The uncertainty is large, but the fraction is significantly higher than any of the other two populations, that it comparatively dominates nevertheless. The remaining part formed from then until $z$$\sim$0.15, apart from the centre which is particular. Here, in the central bin, $\sim$30\% is composed of the intermediate population and only $\sim$60\% of the old population.  

NGC\,6753 demonstrates subtle differences in the M- and L-weighted results. For both, old stars dominate in the centre,
but less significantly in the L-weighted values. We cannot detect any young component in either. At the expected radii of the inner ring, we do not find any major difference in young or intermediate populations. Instead we measure a radially increasing amount of intermediate populations in mass from the centre which stabilizes around 10 arcsec, and remains high ($\sim$40\% in mass). Already in the color profile obtained by \citet{2011Li}, no major indication of this ring was found. In Fig.~\ref{SM-Lum-profiles} we clearly see that the light in the ring regions originates from the intermediate component. Hence this ring has formed between $z$$\sim$2 and $z$$\sim$0.15 or maybe even earlier, like other fossil rings that have been found \citep{2001Erwin}. Furthermore, the fact that we do not detect a single enhanced region corresponding to the current position of the ring might mean that flocculent spiral structure has been present continuously throughout the last few Gyrs and thus has caused this uniform distribution of intermediate populations. Similar to NGC\,5701, a significant part of this galaxy ($\sim$50-70\%) has already been in place at $z$$\sim$2. 

NGC\,7552 shows a dominant young population in the centre, contributing $\sim$50\% to the luminosity-weighted
values. This central starburst, related to the circumnuclear ring, was reported and investigated numerous times in the literature \citep[e.g.][]{Schinnerer1997, Pan2013}. The light of the underlying old population is only dominant in a radius between $\sim$6 and $\sim$10 arcsec, corresponding once more to the region of the high velocity dispersion ring. The mass-weighted contribution is significant over the entire FoV (up to $\sim$75\%), along with the intermediate component (around $\sim$45\%). In these M-weighted results, we barely detect a contribution of the young population dominating the light.  Along the ring of high stellar velocity dispersion (see Fig.\ref{fig:kins}, outside the circumnuclear ring), intermediate and especially old stars are the most significant. We might see an older component, corresponding to the higher velocity dispersion, whose central parts cannot be detected in light (lower $\sigma$) due to the dominant star burst. We report once more the existence of at least $\sim$50\% to $\sim$60\% of the galaxy's mass already at $z$$\sim$2. We also wish to note that in this figure, the main trends are nicely revealed. However, the complexity of spiral arms, starburst, dust lanes and rich substructure probably require an even higher spatial resolution to better understand their interplay.

Our finding of a dominant old component of stars older than 10 Gyrs in all
bulges is consistent with studies on the Milky Way (MW) bulge. Here, colour-magnitude diagrams and spectroscopic studies reveal
that the majority of bulge stars are older than 10 Gyrs
\citep[e.g.][]{1995Ortolani, 2006Zoccali, 2008Clarkson}. 
The metal rich component of the bulge \citep[e.g][]{2010Babusiaux, 2011Johnson, 2012Ness}, shows evidence for comprising stars with a range of ages \citep{2013Bensby}. This metal rich component
exhibits bar-like kinematics and could be associated with the secularly evolved
MW bar and compared to the younger components that we find in NGC\,5701 and NGC\,7552. Similar results
were also found in other external galaxies where the old population dominates in
mass and a younger population would make up only $\sim$25\%
\citep[e.g.][]{macarthur2009}. 

In all galaxies, we can see in the M-weighted results, that the mass gain in
2 < $z$ < 4 (difference of continuous and dashed red line) appears to be rather homogeneous radially, while
the green curve resembles the distribution of young ages in
the L-weighted values. This might be a hint to redistribution of
material over a longer time, i.e. the slightly older stars are by now
well-distributed, while the intermediate component has formed according to similar
processes as the youngest component is forming now. Especially in the two barred
galaxies, it is evident to see that in the last few Gyrs (0.1 <
$z$ < 2), a central component has formed. In both cases, this causes a rise of
up to $\sim$40\% of the central mass. The barred structure could be responsible
for driving the necessary fuel towards the centres of these galaxies in order to
aliment this star formation episode. From theoretical studies we know that bars are
able to remove angular momentum from the gas, driving it to the centre and
enabling star formation \citep[e.g.][]{1981CombesSanders}. We cannot identify such a central structure in NGC\,6753, the
unbarred galaxy in our sample. The absence of this clear central component in the 0.1 < $z$ < 2 population supports the above scenario of the bar influence in the other two galaxies. We also detect a stronger and more confined central increase in intermediate populations for NGC\,5701 than for NGC\,7552. This could be an indicator that in NGC\,5701, secular processes have already started earlier, in concordance with this galaxy being an earlier type.

It is common practice to calculate the Mg over Fe abundance ratio as a measure of the formation time scales of stellar populations \citep[e.g.][]{2003ThomasMB}. It would be optimal to add this parameter to the models which are used for full-spectral fitting. However, they are not yet available. We therefore calculate it from the index-index diagrams, as it is widely done in the literature. As shown before,
this value is only representative when dealing with one single prominent
population. In Fig.~\ref{SFHs}, we detect in most cases a mixture of populations. We decided to determine
the Mg over Fe abundance ratio only for NGC\,5701 and the center of NGC\,6753 based on a dominant fraction of one population (please see Appendix~\ref{alphaE} for illustrations and more details). The central values result to be very high for both,
$\approx$0.25 for NGC\,5701 and $\approx$0.2 for NGC\,6753, indicating a very rapid formation in both cases. Towards the edges of the FoV, this value decreases. This scenario is
consistent with the inside-out-growth model for galaxies where the inner parts
formed before and faster than the outer parts. It also fits the picture of
spatially preserved downsizing presented by e.g. \citet{2013Perez}, stating that the
inner regions in more massive galaxies grow faster than the outer ones. So far, former studies found increased metallicity
and lower [$\alpha$/Fe] values in the central parts of external, but also in the
Milky Way bulge \citep[e.g.][]{2007Jablonka, 2006MoorthyHoltzmann}, while others
detect a variety of different gradients \citep{macarthur2009}. In our study we
find elevated central [$\alpha$/Fe] in combination with a high metallicity.

%-----------------------------------------------------------------------------
\begin{table*}
\centering
\caption{Light and mass fractions in \% in the central (< 0.3 kpc) and inner (0.5 kpc < r < 2 kpc) parts of young (<\,1.5\,Gyr, redshift $z$ < 0.1) intermediate (1.5\,Gyr\,<\,intermediate\,<\,10\,Gyr, 0.1 < $z$ < 2) and old (>\,10\,Gyr, $z$ > 2) populations. Regions are indicated in Fig.~\ref{SFHs}.}
\begin{tabular}{llclrcr}
\hline
Galaxy & ~  & centre (< 0.3 kpc) & ~ & ~ & inner (0.5 kpc < r < 2 kpc) & ~ \\
 ~ & young & intermediate & old & young & intermediate & old \\  
 \hline
NGC\,5701 - L-weights & 0.15$\pm$0.09  &  36.6$\pm$3.2   &  63.2$\pm$8.8  & 0.0$\pm$0.0  &  38.9$\pm$8.4   &  61.1$\pm$9.9  \\
NGC\,6753 - L-weights & 0.30$\pm$0.01  &  42.5$\pm$0.2   &  57.2$\pm$0.2  & 1.8$\pm$0.9  &  49.2$\pm$3.1   &  49.0$\pm$3.7  \\
NGC\,7552  - L-weights& 48.5$\pm$0.5  &  33.4$\pm$1.7   &  18.1$\pm$1.3  & 14.0$\pm$4.7  &  44.6$\pm$6.6   &  41.4$\pm$8.5  \\
\hline
NGC\,5701 - M-weights & 0.01$\pm$0.01  &  30.6$\pm$4.8   &  69.4$\pm$4.9  & 0.0$\pm$0.0  &  14.1$\pm$3.2   &  85.9$\pm$3.3  \\
NGC\,6753 - M-weights & 0.03$\pm$0.02  &  28.0$\pm$0.2   &  72.0$\pm$0.7  & 0.2$\pm$0.2  &  44.7$\pm$3.8   &  55.1$\pm$4.0  \\
NGC\,7552  - M-weights& 4.1$\pm$0.1  &  43.4$\pm$1.7   &  52.5$\pm$1.9  & 0.6$\pm$0.4  &  28.0$\pm$8.0   &  71.4$\pm$8.1  \\
\hline
%\footnotesize{NOTES: }
\end{tabular}
\label{tab2}
\end{table*}
%-----------------------------------------------------------------------------

\subsection{Implications for bulge evolution models}
\label{results3}
Despite the small number of galaxies investigated, our analysis allows to help putting constraints on theoretical models, trying to understand the build-up and evolution of galactic bulges as: 1) we selected a representative of early and late-type spirals which seem to exhibit the typical characteristics (see Fig.~\ref{fig:ind-dis}) and 2) we find common results within the three galaxies hinting towards a similar origin and fundamental evolution process, across these types.

In particular, we tried to get a handle on radial
stellar mass distributions in the present day Universe. Despite the differences of the three investigated bulges, we find a
significant amount of old stars at all radii. Hence,
at least 50\% of the stellar mass was already formed at $z = 2$ and even $z$ = 4 (with increasing
percentage from NGC\,7552 to NGC\,5701 where we find more than 80\%). Furthermore, we detect a significant fraction of mass in a second star formation episode below $z$=2. Its present day distribution is more localized and can be associated with current morphological features such as bars.

A wealth of numerical models have already explored the early formation of galaxies and their central components. The main catalysts for the first stellar formation periods have been identified as mergers \citep[e.g.][]{1992Hernquist, 2005Bournaud,2010Hopkins}, the collapse itself, e.g. in the $\Lambda$CDM \citep{1978WhiteRees}, or high-$z$ starbursts \citep[e.g.][]{2013Okamoto, 2013Finkelstein}. Independent of the model used, the maximal percentage of bulge mass formed before $z$=2 is usually no more than 50\% and often less (ranging from 10\% to 50\%, see also \citealt{2013Obreja}). Thus, the remaining mass percentage is supposed to be attributed to later evolutionary processes, related to a second star formation peak between redshift 1 and 2 \citep[e.g.][]{1996Madau, 1997NormanSpaans, 1997SpaansCarollo,2010Daddi}. Nonetheless, these processes might still be one of the above, but due to lower mass densities (expansion of the universe), they are more likely of secular origin. In particular when features can be associated to morphological structures such as bars, rings, (nuclear) spiral structure etc., the likelihood of internal (and/or secular) evolution increases. 

In all our galaxies we detect these different components, but we always find a higher percentage of old stellar mass than found in simulations. In the very central parts (r<0.3 kpc), the old population comprises above 50\% in one and around 70\% in two out of the three bulges (see Tab.~\ref{tab2}). This percentage is on average even higher (up to $\sim$85\% ) considering the inner parts (0.5 kpc < r < 2 kpc). Hence, whichever process(es) were acting in the early life of our three galaxies, they must have produced more stellar mass as commonly predicted. This is however only revealed by recovering their mass-weighted results. 

All galaxies also display regions of enhanced intermediate (and young) populations which can be associated to morphological structures. Therefore, we suggest that these populations are related to a secular origin. In the following we will briefly discuss each galaxy and speculate about their formation scenario, based on our results and model comparisons.

Earlier studies already report a strong influence of
environment on the resulting bulge types \citep[e.g.][]{2009Kormendy}. In
high density environments mergers are more probable to occur and influence the bulge formation, leading to old elliptical-like structures. None of the investigated galaxies shows signs of recent interaction and do not
have close neighbours. NGC\,5701 forms part of the Virgo supercluster as a
member of the Virgo III Groups. Hence, it could have suffered mergers more likely than the other two galaxies leading to the highest percentage of old stellar mass of the three, both in inner and central regions. 

Both, NGC~6753 and NGC~7552 show similar percentages but different distributions of old and intermediate populations. Their morphologies could be key for this. In NGC\,6753 no bar is
present and we can distinguish much better the centre composed of old stars,
high in L- and M-weights. Along with its kinematic properties (high stellar velocity dispersion), the centre of this galaxy could present the relict of a node where the first SF occurred \citep[e.g.][]{2013Obreja, 2013Barro}

Both NGC\,5701 and NGC\,7552 host large-scale
bars which have affected the populations formed between z$\sim$2 and z$\sim$0.1,
during the major second star formation epoch as predicted by simulations and found observationally. Here, the
central parts show an increase of these intermediate stars: between 35-50\% of the
mass fraction. They can be attributed to the influence of the
bar affecting this population (formed between z$\sim$2 and z$\sim$0.1) in particular.

We speculate that all three galaxies thus suffered a common initial stage of collapse (and/or early SF), but while NGC~5701 may have been affected also by its denser environment, the other two could retain material to form more stars in later epochs. Hence, the contribution of old stars dominating in NGC~5701 might lead to the photometric classification of a ''classical bulge'' \citep{weinzirl2009}, while intermediate (and young) populations are almost equally important in NGC~6753 and NGC~7552, likely resulting in a photometric  ''pseudo-bulge'' classfication in the same former work.

%###############################################################################
\section{Summary and Conclusions}
\label{conclusions}
In this paper we present kinematic and stellar population maps of three
significantly distinct bulges with the aim to quantify the importance of different populations to better constrain their evolutionary scenarios. Our data differ from most current integral field
surveys, since we obtain a very high spatial (elements of $1''$x$1''$) as well
as -
and especially - high spectral resolution of R$\approx$7000. This combination
allows
us to study the inner regions in galaxies in great detail and carry out a
comprehensive stellar population analysis using the full-spectral fitting code {\tt STECKMAP} \citep[e.g.][]{2006MNRAS.365...74O,  2006MNRAS.365...46O, 2008Kolevab, 2011MNRAS.415..709S}. We employ a novel analysis method interpreting its 2D results by deriving different stellar components and their
contributions to the overall mass and light profiles of the galaxy. In addition we point out clearly
the
limitations that a mixture of populations can cause in deriving abundance
ratios using classical line-strength methods.
\\ \\
Analyzing the kinematics and stellar populations of the three bulges, we deduce
different formation scenarios:
\begin{itemize} 
\item NGC\,5701 consists of mainly one old solar (or slightly sub solar) stellar
population, where up to $\sim$80\% of the galaxy's mass already existed at $z$$\sim$4
distributed now radially almost uniformly (in our FoV), apart from the centre. The [Mg/Fe] values confirm a fast origin. The extremely weak nuclear
spiral can be detected in the higher order moment maps ($h3$, $h4$) and in the stellar light contribution. Despite its stellar bar, it does not exhibit additional star formation, but shows a significant amount of intermediate populations in the central 0.5 kpc. It supports inside-out growth
and appears to show a classical bulge (consistent with previous photometric analyses and increased $\sigma$)
which formed almost simultaneously along with the bar.
\item NGC\,6753 falls into three regions: the centre is old, metal-rich, with
high
[Mg/Fe] values, and an extremely high stellar velocity dispersion while outside of it, $\sigma$ drops and we find the presence of a significant intermediate population. We also detect an inner ring in ionised gas emission, $h_4$ moment and light of younger/intermediate populations. Yet, the main stellar
ages
vary and different populations are distributed throughout our FoV,
suggesting the presence of former, short lasting ring or spiral structures
producing
the wide age range of intermediate (above 1.5 Gyrs) stars. Nevertheless the main
component in mass is composed of old (above 12 Gyrs) stars at all radii
($\sim$50-60\%). 
\item NGC\,7552 shows three regions: the starburst centre/circumnuclear ring, a high stellar velocity dispersion ring
and
an underlying disk component. The centre is dominated by the starburst and shows a
young stellar component with around solar to sub-solar metallicities.
 At
the inner side of the ring a metal-poor, old component can be found superimposed
to a
younger component. The outer parts of the ring present a smooth trend towards the
intermediate-to-old solar stars plus younger and less metal-rich that we find in
the
rest of the galaxy. The central young component,
extremely
dominant in light, almost does not show in the mass-weighted results
demonstrating
its rather recent formation. Here a significant fraction
($\sim$50-60\%) of the stellar mass formed again before $z$$\sim$4.
\end{itemize}

We find in all cases that most of the stellar mass has been formed long ago (before z$\sim$4) -
with a tendency to decrease for later types. We also find a strong influence of
the bar on the stellar component formed between z$\sim$2 and z$\sim$0.1. 

Comparing our results with specific simulations and models, we can confirm a two-fold formation
process of galactic bulges as suggested by e.g. \citet{2013Obreja}: a rapid formation
of an old bulge structure in the early cosmic web initiating star formation in
dense nodes (with possible influence of mergers, at least in NGC~5701) and a slower formation during the high star formation period between
redshifts $\sim$2 and $\sim$1 of a younger component.
Our results do not agree with the simulations on the mass fractions found. In all our cases the mass fraction of the old stellar component is larger - more than 50\% and up to 80\% - than the predicted $\sim$30\% or maximally $\sim$50\%, pointing to higher star formation
efficiencies or distinct evolutionary processes in the past. The secularly evolved component (here: intermediate populations) account for 30-40\%. 

The limitations of our sample not only concerning its size, but also its mass range, are obvious and it cannot be regarded as representative. Nevertheless, our results point towards a common conclusion, namely the formation of already higher stellar mass fractions in the early universe than currently predicted. This work calls for even more detailed studies quantifying the importance of each
process - collapse, starbursts, mergers and secular - at a given point of the lifetime of a
galaxy in order to fully understand its evolutionary path. We will try to take a
second step in Cacho et al. (in prep), where we extend this work to separate the
kinematics associated to each population in our sample of galaxies.

\section*{Acknowledgments}

We thank Alexandre Vazdekis for helpful suggestions, Tim de Zeeuw for a critical reading of the original version of the paper and an anonymous referee for very useful comments. MKS, RC and JFB wish to express their gratitude to the Mount Stromlo Observatory and researchers for their friendliness and support during our extended stay. MKS acknowledges the support of the Instituto de Astrof\'isica de Canarias via an Astrophysicist Resident fellowship and Ignacio Mart\'in-Navarro, Andra Stroe and St\'ephane Courteau for useful discussions. RC acknowledges the Ministerio de Ciencia e Innovaci\'on by means of their FPI program (grant AYA-2010-21322-C03-03 and AYA-2013-48226-C3-3-P). JFB acknowledges support from the Ram\'on y Cajal Program and from the FP7 Marie Curie Actions of the European Commission, via the Initial Training Network DAGAL under REA grant agreement number 289313. TRL thanks the support of the Spanish Ministerio de Educaci\'on, Cultura y Deporte by means of the FPU fellowship. This research has been supported by the Spanish Ministry of Economy and Competitiveness (MINECO; grants AYA2010-21322-C03-02 and AYA2009-11137) and by the Spanish Ministry of Science and Innovation (MICINN; grants AYA2011-24728 and Consolider-Ingenio CSD2010-00064) and by the Junta de Andaluc\'ia (FQM-108).

\bibliographystyle{mn2e_fixed.bst}
\bibliography{refs_jabref_v2}

\newpage

\appendix{}
\section{Galaxies}
\label{app:galaxies}
\textbf{NGC\,5701} is an early-type galaxy with a rather smooth bulge. According
to
the De Vaucouleurs Atlas Description, it shows a well-defined bar imbedded in a
strong inner lens, one of the best-defined examples of this phenomenon. In the
centre and bar region, this galaxy does not seem to exhibit star formation nor
dust.
According to \citet{2002Erwin}, this galaxy (being part of the WIYN Sample) has
no
inner structures apart from a nuclear spiral. Therefore, here we expect to have
one
dominant old population with possibly a weak younger population in the centre.
Furthermore, we can test the hypothesis brought forward by
\citet{2003GadottideSouza} to be a disk-lacking galaxy. 

\textbf{NGC\,6753} is also an early-type galaxy but with more substructure. The
dominant feature here is a bright inner ring which lies at the rim of a fairly
uniform inner disk zone. This inner disk zone is filled with H-alpha emission
\citep{1996CrockerBB}. Outside the inner ring, a broad oval zone includes
complex
and partially flocculent spiral structure. Beyond the broad oval zone, a
well-defined, and mostly detached outer ring is found. For this project, mostly
the
inner parts are of importance and here we now expect to find more substructures
than
in NGC\,5701. In particular, we aim to detect signatures of an inner ring both
in
the kinematics and stellar population parameters.

\textbf{NGC\,7552} is mostly defined by a complex, dusty bar and is best known
for
its central starburst, which is associated with a nuclear ring. The bar is
prominent, and numerous H{\sc ii} regions are scattered within the disk in an
asymmetric pattern. The 1\,kpc starburst ring is best visible in radio
wavelengths
and reveals numerous supernova remnants \citep{1994Forbesa, 1994Forbesb}.
Nevertheless, it does not present very strong nuclear activity which simplifies
studies of the circumnuclear ring. The classification of this object seemed
difficult amongst the literature: The presence of giant H{\sc ii} regions near
the
corotation radius lead \citet{1989Bonatto} to the conclusion to deal with an H
II
galaxy, whereas \citet{1988Durret} classified it as a LINER due to the detection
of
a weak [O{\sc i}] $\lambda$ 6300 line. The dusty bar morphology is very unusual
(the
De Vaucouleurs Atlas Description). As a member of the Grus triplet, the galaxy
may
be affected by an interaction that has disturbed its morphology. In fact,
\citet{1992Claussen} report high molecular gas concentrations in the centre and
signatures of tidal disturbance deduced from the observed asymmetries in  their
CO
line profiles. Additionally, \citet{1990Feinstein} discovered two weaker rings
(of
radii 1.9 kpc and 3.4 kpc). \citet{1999HameedD} investigate NGC 7552 via
H$\alpha$,
while \citet{1994Forbesb} focuse on the ring, revealing yet another inner ring
of
the size of only 1 kpc in the radio. \citet{Schinnerer1997} concentrate as well
on
this central feature showing among others Br$\gamma$ images, also tracing gas
ionized by recently formed massive stars and distinguish different star
formation
histories (SFH) for the nucleus and the ring. Furthermore, based on NIR and HST
V-band continuum maps, they postulate the existence of in inner bar located
inside
the nuclear ring and perpendicular to the outer east-west oriented large bar. 
\citet{Pan2013} discuss in detail the circumnuclear starburst ring and the
related
formation of dense molecular gas and stars in that region.

%\section{Model testing as limits of LS analysis: example NGC\,6753}
\section{Model testing as limits of LS analysis: example NGC~6753}
\label{milesmodtest}

%-----------------------------------------------------------------------------
\begin{figure*}
\includegraphics[width=1.\linewidth]{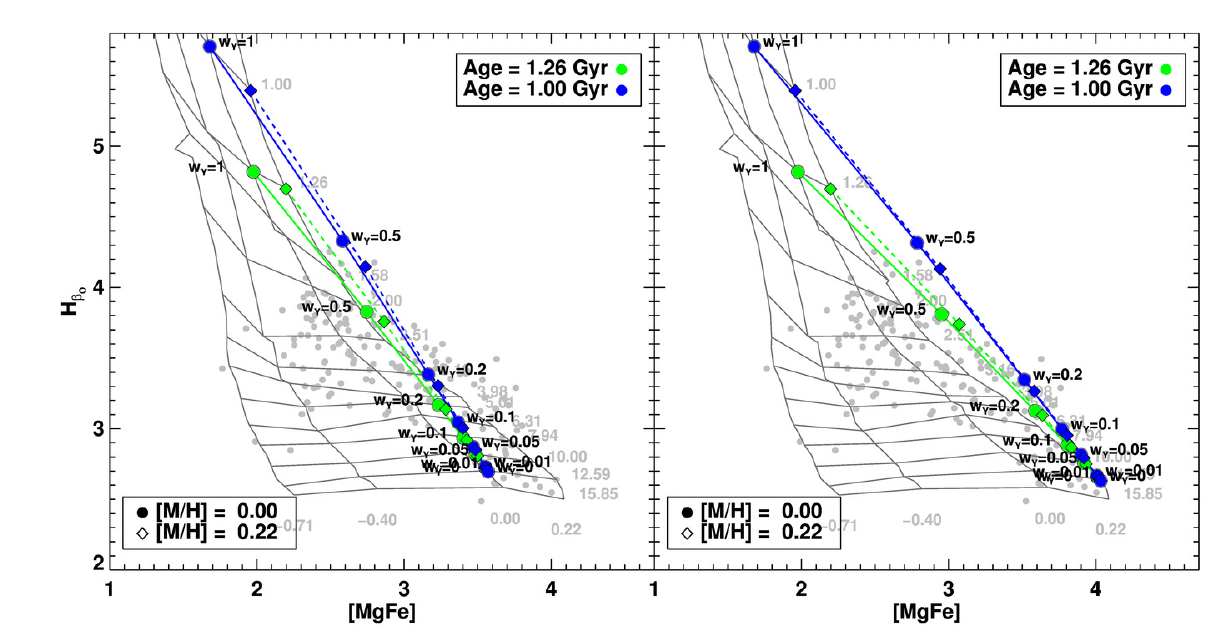}
\caption{The influence of the combination of very different sets of stellar
populations on the index results. We show model tests in colors: green and
blue  points indicate age, round symbols solar metallicity and rhombic symbols
super  solar metallicity. The different fractions of young populations are
indicated next to the points, for the 1.00 Gyr population on the right, the
1.26 Gyr, on the left. In addition, we display the index measurements as
obtained from NGC\,6753 in grey dots. See section~\ref{milesmodtest} for more
details.}
\label{fig:miles-test-lum}
\end{figure*}
%-----------------------------------------------------------------------------

In order to explain the points outside the grid for NGC\,6753 seen in Fig.~\ref{fig:5701-grid}, we combined
different
single stellar population models, changing their weights, ages and
metallicities, similar to Fig.~8 in \citet{Kuntschner2000}. We
use the same SSP MILES models which we use for the SSP
grid\footnote{http://miles.iac.es/pages/webtools/get-spectra-for-a-sfh.php}. 
From
the individual spectra for a certain age and metallicity for each population, we
create a final spectrum which we analyze using the same line-strengths indices
routine which we apply to the galaxy spectra. When combining the spectra, we
impose
the contribution in light per population. In the two test cases we show here, we
chose two different metallicities for the young population: 0.00 and 0.22 dex,
and
two ages: 1.00~Gyr and 1.26~Gyrs, and the following weighting scheme:

\begin{itemize}
\item{100\% young stars}
\item{50\% young and 50\% old stars}
\item{20\% young and 80\% old stars}
\item{10\% young and 90\% old stars}
\item{5\% young and 95\% old stars}
\item{1\% young and 99\% old stars}
\item{100\% old stars}
\end{itemize} 

Fig.~\ref{fig:miles-test-lum} summarizes the test outcome when combining an old
($\approx$ 12.5 Gyr) population with solar metallicity (left column) with the
two
different young populations of each two different metallicities. On the right
plot,
we show the same, just with the older population having super solar metallicity.
 
We also performed this same analysis weighing the spectra in mass. Here, it was
much
easier to move points outside the grid. Already very small mass fractions of a
young
population resulted in a point outside in the measurement of the combination.
This
is due to the fact that even a small fraction in mass of a young population ($\sim$1-5\%, depending on the exact age and its metallicity) will
have
a strong contribution in light and therefore outshine most of the old
population.
Since the index measurements are based on luminosity-weighted quantities, we
foster
our analysis with the L-weighted tests. 

Based on these tests, we conclude that NGC\,6753 presents an old, metal-rich 
(about solar) population, significant in mass, but whose light is mixed with a strong younger population. The old
population is seen in the inner parts. Slightly further out, the young
population
begins to contribute more in light; according to the tests, we need
approximately
20\% of the light contribution coming from the young population (and only 1 to
5\%
in mass). Therefore, the index values that we measure move upwards and hence out
of
the grid. Thus, these values do not indicate a failure of our measurements, but
the
combination of a rather metal-rich population in combination with a low
mass-fraction of young stars. 

More importantly, this result implies that abundance
ratios
measured in the region of population mixtures will be unreliable if it is sufficiently altered by the above effect.

%-----------------------------------------------------------------------------

\section{$\alpha$ - enhancement}
\label{alphaE}

 %-----------------------------------------------------------------------------
\begin{figure*}
  \includegraphics[width=1.\linewidth]{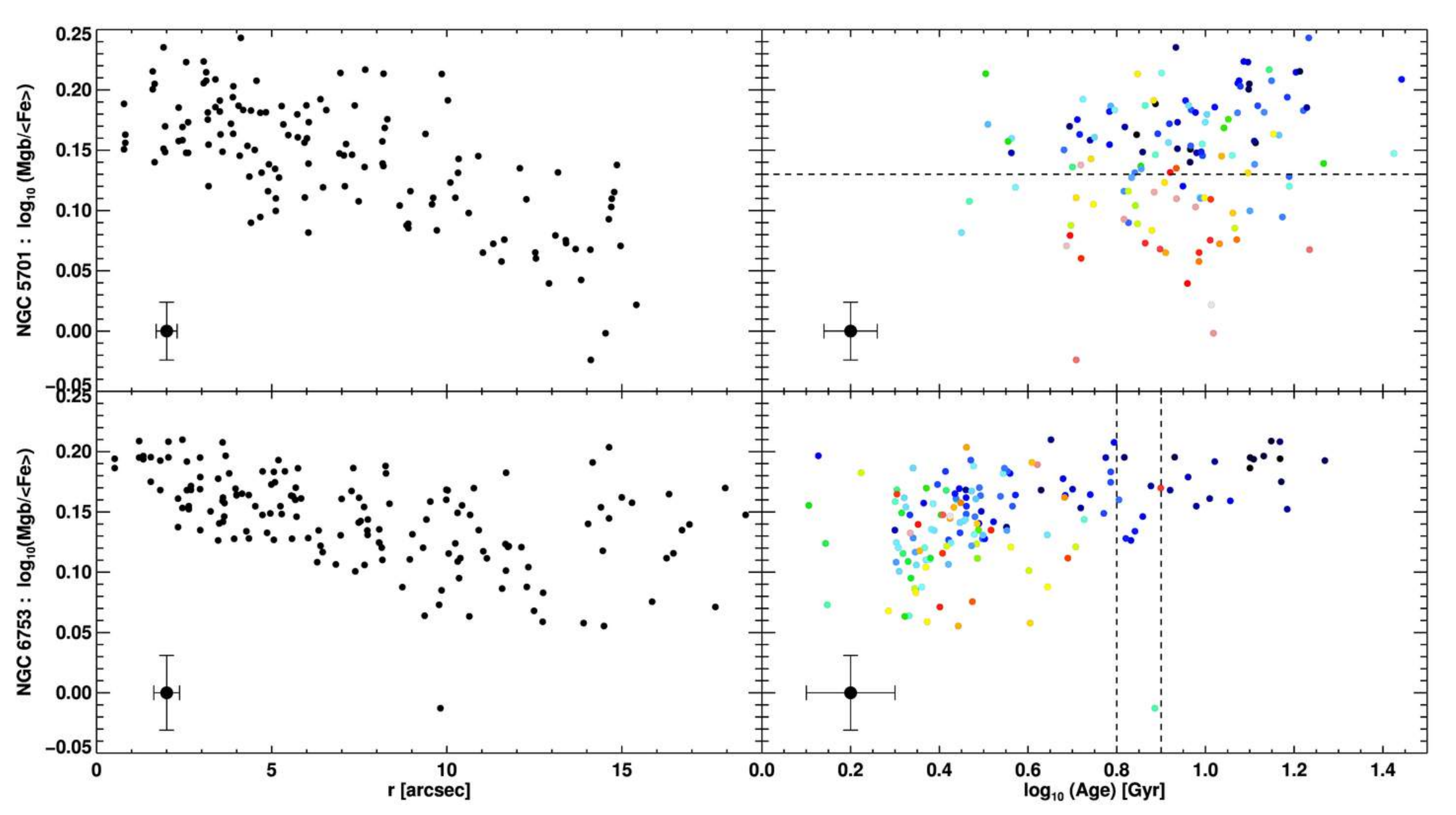}
  \caption{We show abundances as a function of radius for and age for two galaxies: upper row: NGC~5701, lower row: NGC~6753. The horizontal line indicates for NGC~5701 the separation of the central bulge and the outer bar dominated region. Vertical lines (drawn at 6.31 and 7.94 Gyrs) separate the region where the values can be trusted (right side) and where, due to an obvious mixture of populations, we cannot trust the values any longer (left side); the region in the middle could be debatable. Representative error bars are given in the left lower corners.}
  \label{abun}
\end{figure*}
%-----------------------------------------------------------------------------

In Fig.~\ref{abun} we present the results of the abundance ratio analysis for two of the three galaxies. On the left hand side, the abundance is plotted as a function of radius indicating the overall decline from the center to the outskirts of the two galaxies, with increasing scatter in the individual values, especially for NGC~6753. On the right, the abundance versus the age is shown with color-coded points in the same way as before: darker points are central ones and yellow, red points belong to the edges of the field. 

NGC~5701 shows a separation of two clouds indicated by the horizontal line. The ages are rather homogeneously distributed. 
The profile of NGC~6753 reveals certain details: going outwards, we can recognize a series of bumps. We deliberately chose to represent individual points here since averaging even in ellipses would wash out signatures of the patchy spiral structure. Comparing this profiles thus with the unsharp mask in Fig.~\ref{fig:sample}, we can qualitatively correlate the wiggles in the profile with the spiral arms. Nevertheless, in this region the values cannot be fully trusted as pointed out before which might actually be the result of the wiggles. On the right, we indicate with the two vertical lines the different regions: to the right, we can trust the points. This narrow regime without too much scatter corresponds to the central part of the galaxy (those points which do not fall out of the grid in Fig.~\ref{fig:5701-grid}). In between the lines, the points start to be less reliable and to the left, we cannot fully trust them due to the mixture of different populations (the increasing scatter also hints to this problem).

\section{Age binning and Steckmap analysis}
\label{app:explots}

%-----------------------------------------------------------------------------
 \begin{figure*}
 \includegraphics[width=1.\linewidth]{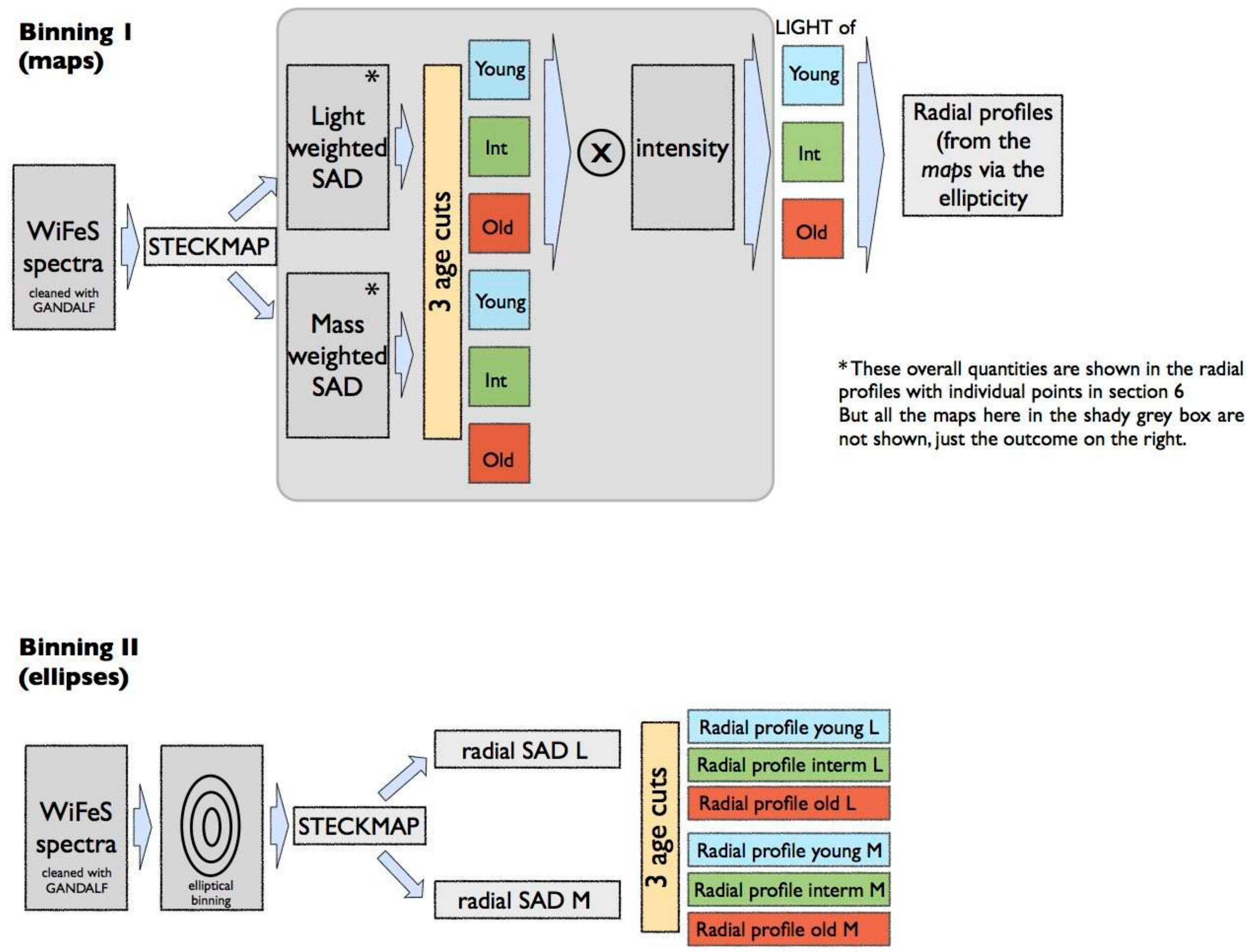}
 \caption{Visualization of the two distinct binning schemes.  \textit{Top:} Binning 1 via maps, where the grey area indicates results that we obtain during the process but do not show in the paper). The corresponding figures using this scheme are Fig.~\ref{fig:SM-TRL} and Fig.~\ref{SM-Lum-profiles}. \textit{Bottom:} Binning 2 via ellipses used for the final part of our analysis, shown in Fig~\ref{SFHs}.}
 \label{Ex1}
 \end{figure*}
%-----------------------------------------------------------------------------

In Fig.~\ref{Ex1} we illustrate the two different processes of using {\tt STECKMAP}, both times departing from the WiFeS spectral cube, using the emission-cleaned spectra coming from the {\tt GANDALF} analysis, shifted
to rest frame according to the stellar velocity (see Sec.~\ref{stkin})
and broadened to 8.4\,\AA. In every case, we fix the stellar kinematics and fit
exclusively for the stellar content avoiding the
metallicity-velocity dispersion degeneracy \citep{2011MNRAS.415..709S}. 

\begin{figure*}
 \includegraphics[width=1.\linewidth]{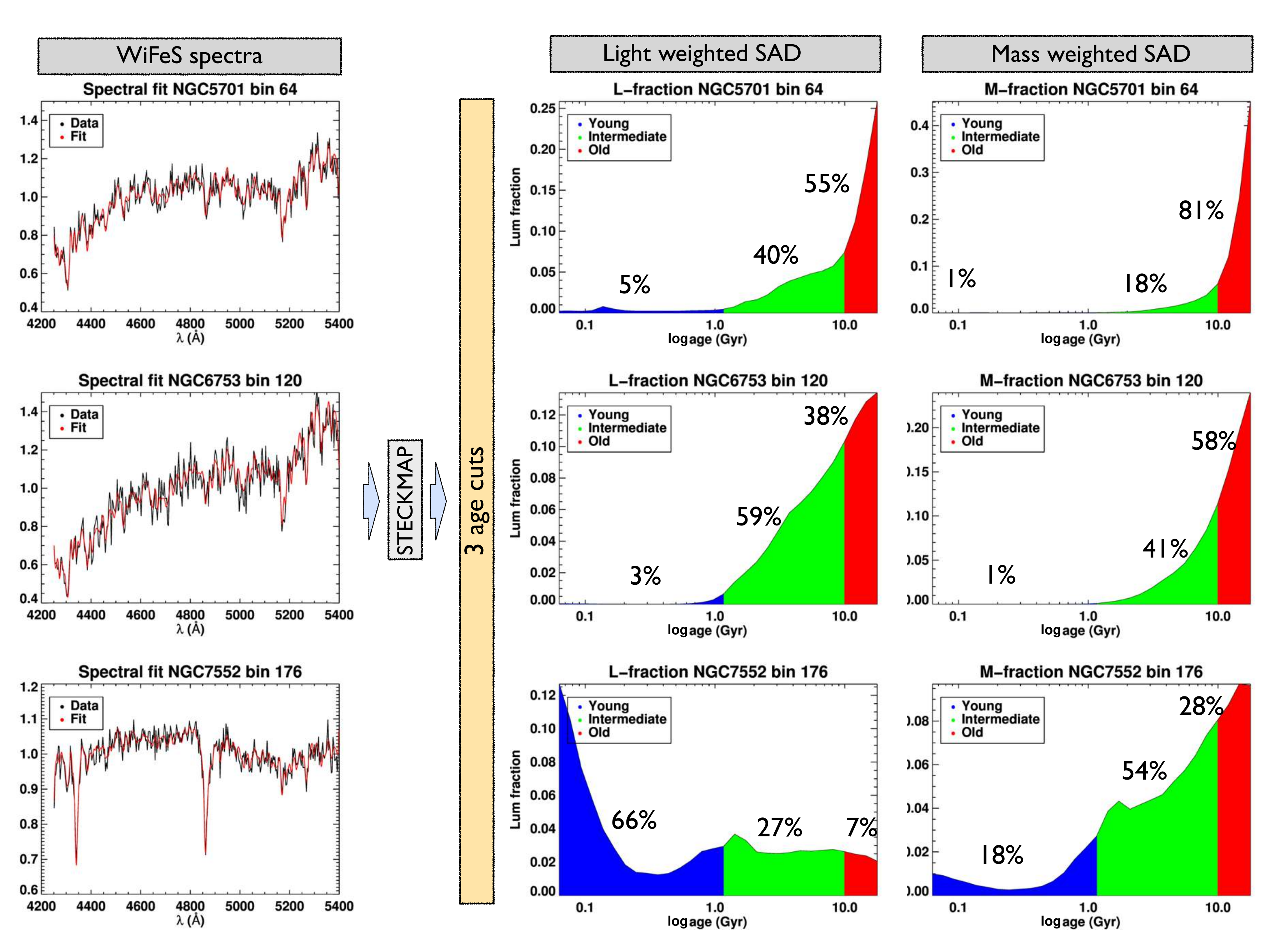}
 \caption{Visualization of the different age bins obtained from the {\tt STECKMAP} SADs. \textit{Left:} Data (black) with the {\tt STECKMAP} fit (red) over plotted. \textit{Right:} The L- and M-weighted SADs with the young (blue), intermediate (green) and old (red) (also from left to right on each plot) fractions indicated (areas under the SAD curve). }
 \label{Ex2}
 \end{figure*}

\textbf{Binning 1 (maps)}: We use this binning, a Voronoi binning over the two-dimensional maps, almost throughout our entire analysis, starting with the kinematics, then the index analysis and later the first analysis with { \tt STECKMAP}. Hence, from these maps we obtain with {\tt STECKMAP} light and mass weighted stellar age distributions (SAD) and from these distributions, we obtain fractions in our three (four) defined age cuts. This process of dividing the SADs into the distinct age bins is visualized in Fig.~\ref{Ex2}. As shown in Fig.~\ref{Ex1}, we do obtain at first corresponding maps for the L- and M-weighted SADs, which we don't display in this paper due to simplicity. From those, we extract the corresponding age fractions, again in maps. Multiplying the L-weighted fractions with the overall intensity, we obtain the light maps corresponding to each of the populations. Using the ellipticity, we then plot the radial profiles directly from those maps, as shown in Fig~\ref{SM-Lum-profiles}. 

\textbf{Binning 2 (ellipses)}: This second binning scheme is only employed in the final analysis in order to raise the S/N. Here we perform an elliptical binning prior to the {\tt STECKMAP} analysis. The radially binned spectra are then analyzed with {\tt STECKMAP} to produce directly radial L- and M-weighted SADs from which we extract once more the fractions of young, intermediate and old populations, as shown in Fig.~\ref{SFHs}. 

 Figure~\ref{Ex2} illustrates for three example spectra (one for each galaxy) the process of obtaining the different age fractions from the SADs. On the left we show the data and {\tt STECKMAP} fit and on the right the corresponding SADs, once L- and once M-weighted. On each plot, we indicate the area under the SAD curve which corresponds to the young, intermediate and old SP. The associated fractions are given on top of these areas.

%-----------------------------------------------------------------------------
%-----------------------------------------------------------------------------

\newpage
\section{Complete maps}
\label{app:maps}
We show maps of the stellar and ionized-gas distribution and kinematics (from the blue grating) as well as indices from the blue and red spectra for the three galaxies of this study. First row: HST or Spitzer image, its unsharp-mask and the name, Hubble type, position, absolute B-band magnitude and inclination of the galaxy. Second row: (i) stellar mean velocity V (in km s$^{-1}$), (ii) stellar velocity dispersion $\sigma$ (in km s$^{-1}$), (iii) and (iv) Gauss-Hermite moments h3 and h4. Third row: (i) H$\beta$ flux (in logarithmic scale), (ii) same for the H$_\gamma$, [O{\sc iii}] and [N{\sc i}] line, (v) mean radial ionised gas velocity and (vi) ionised gas velocity dispersion (in  km s$^{-1}$ ). Fourth row: index maps from the blue spectra for (i) H$\beta$, (ii) H$\beta$$_o$, (iii) Fe5015, (iv) Mg$b$, (v) Ca4227 and (vi) G4300. Fifth row: index maps from the blue spectra for (i) Fe4383, (ii) Fe4668, (iii) Ca4455, (iv) Fe5270, (v) Fe5335 and (vi) Fe5406. Sixth row: index maps from the red spectra for (i) CaT, (ii) CaT*, (iii) PaT, (iv) Ca1, (v) Ca2, (vi) Ca3 - as defined in \citet{2001Cenarro1}. The cut levels are indicated in a box on the right-hand side of each map. 

%-----------------------------------------------------------------------------
 \begin{figure*}
 \includegraphics[width=.65\linewidth]{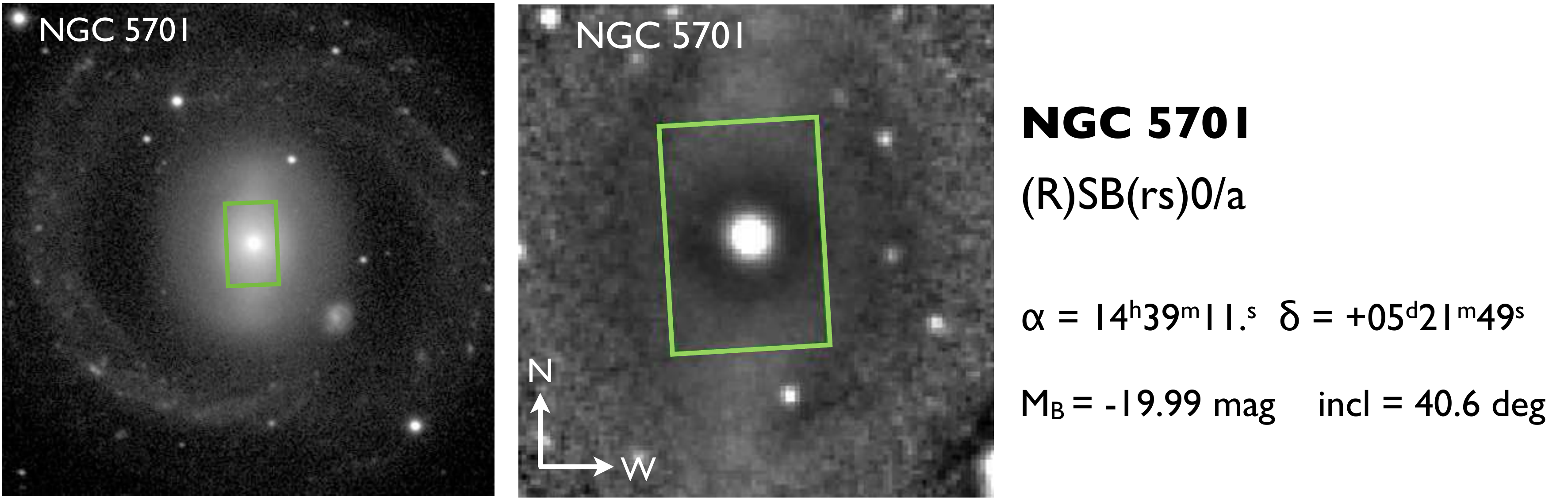}
 \includegraphics[width=.9\linewidth]{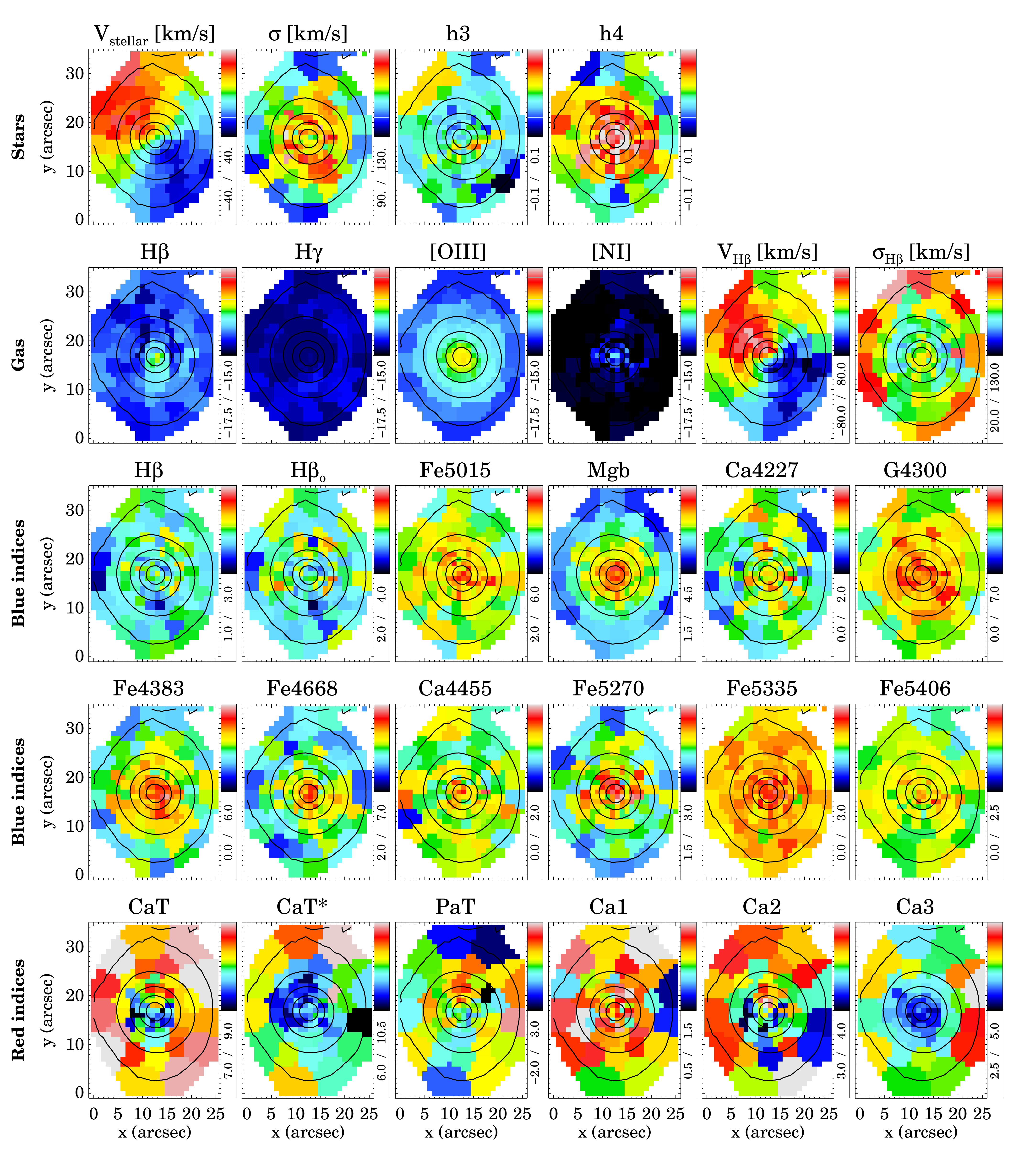}
 \caption{Summarized maps for NGC\,5701 from blue and red gratings. See text for details. }
 \label{5701all}
 \end{figure*}
%-----------------------------------------------------------------------------

%-----------------------------------------------------------------------------
 \begin{figure*}
 \includegraphics[width=.65\linewidth]{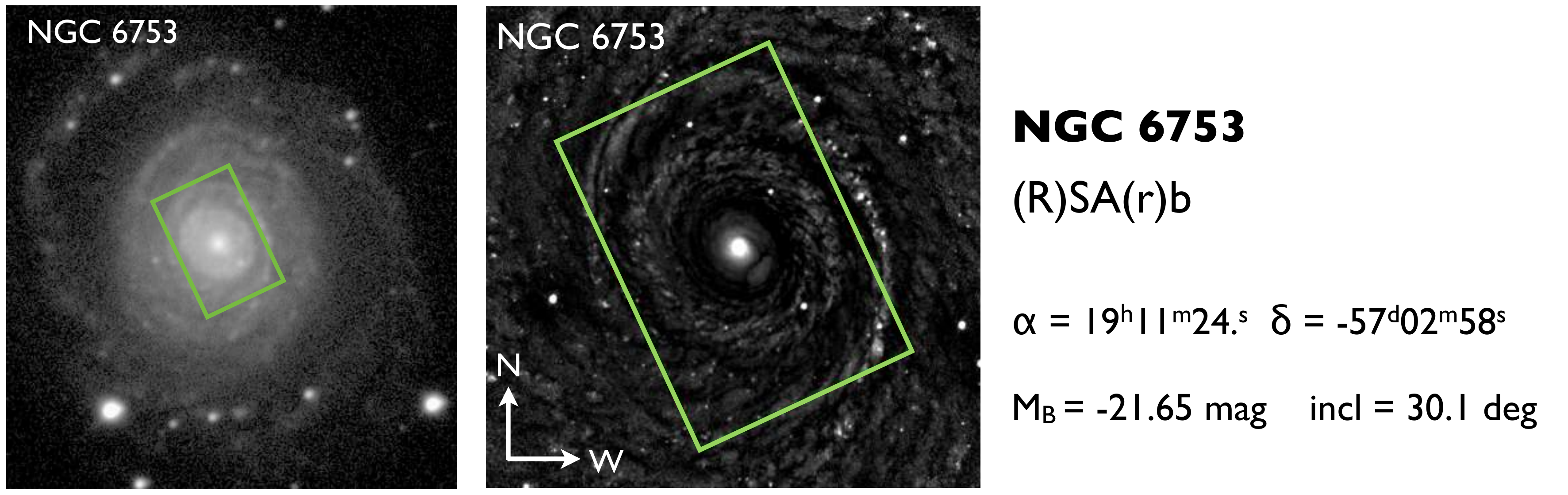}
 \includegraphics[width=.9\linewidth]{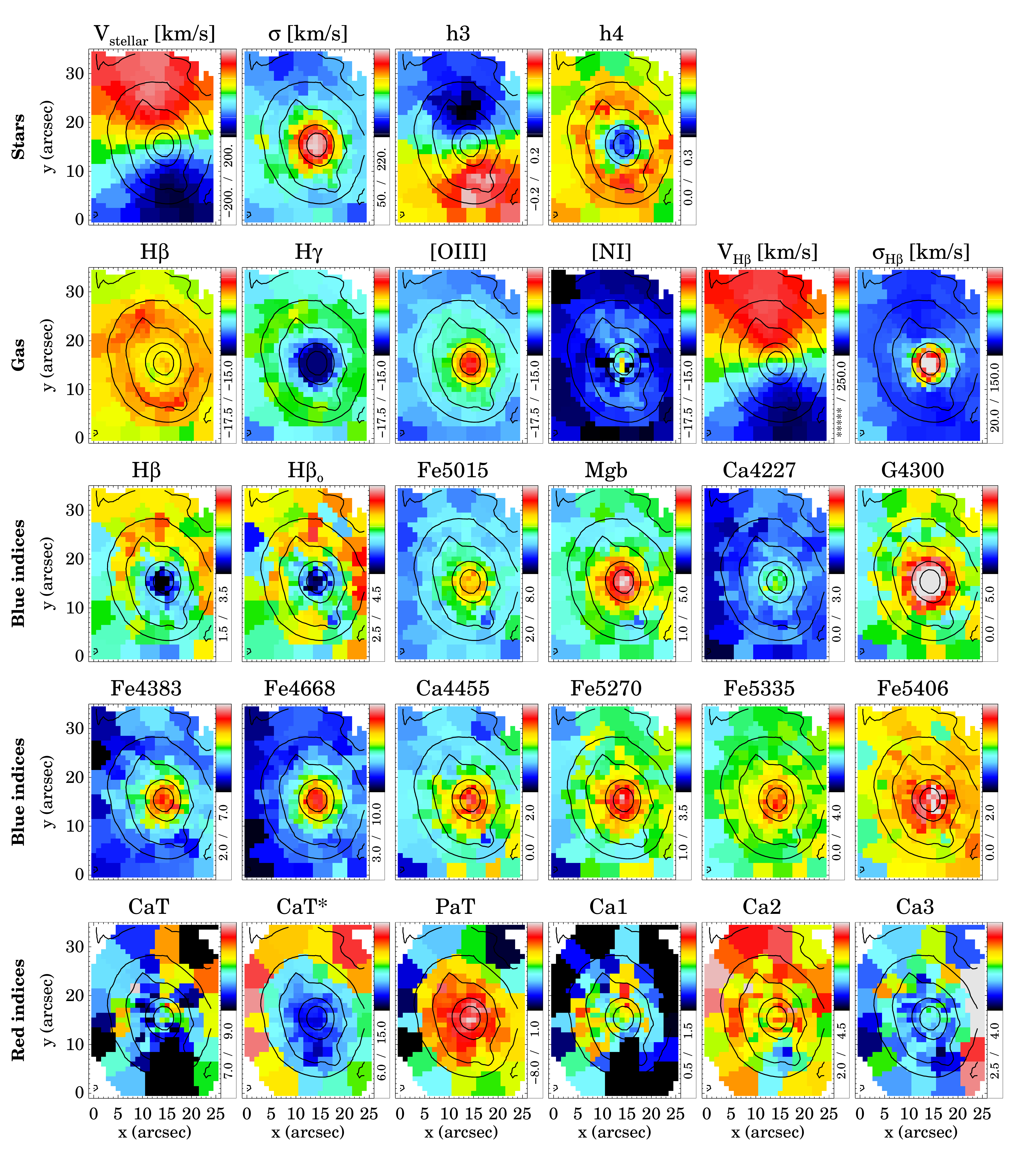}
 \caption{Summarized maps for NGC\,6753 from blue and red gratings. See text for details. }
 \label{6753all}
 \end{figure*}
%-----------------------------------------------------------------------------

%-----------------------------------------------------------------------------
 \begin{figure*}
 \includegraphics[width=.65\linewidth]{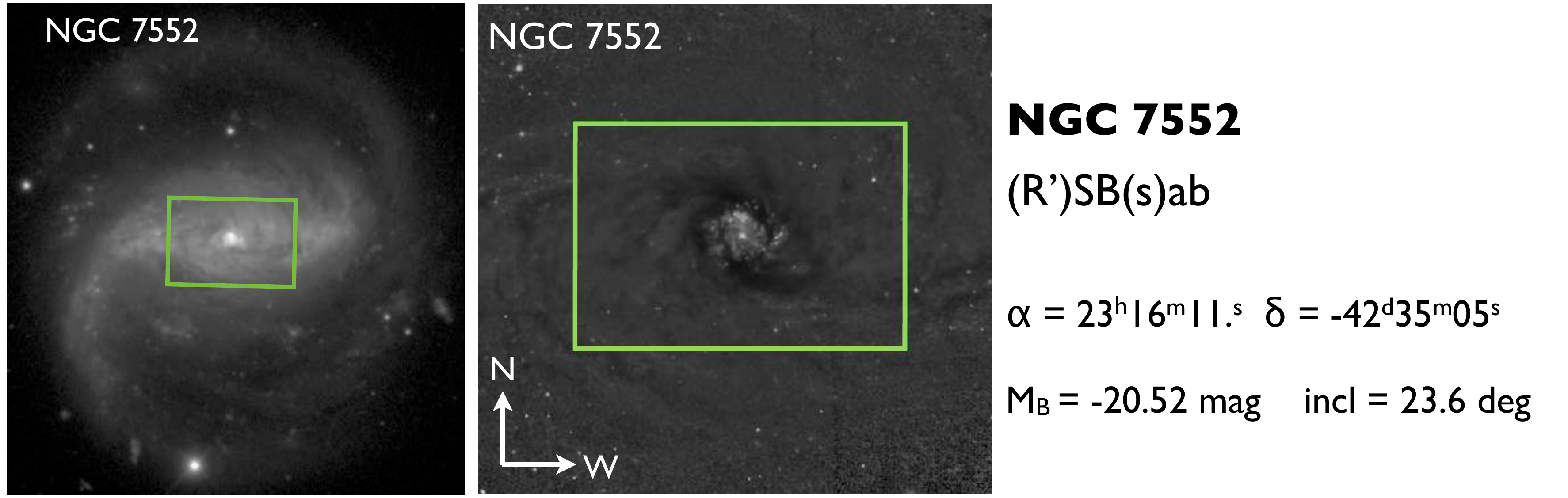}
 \includegraphics[width=.9\linewidth]{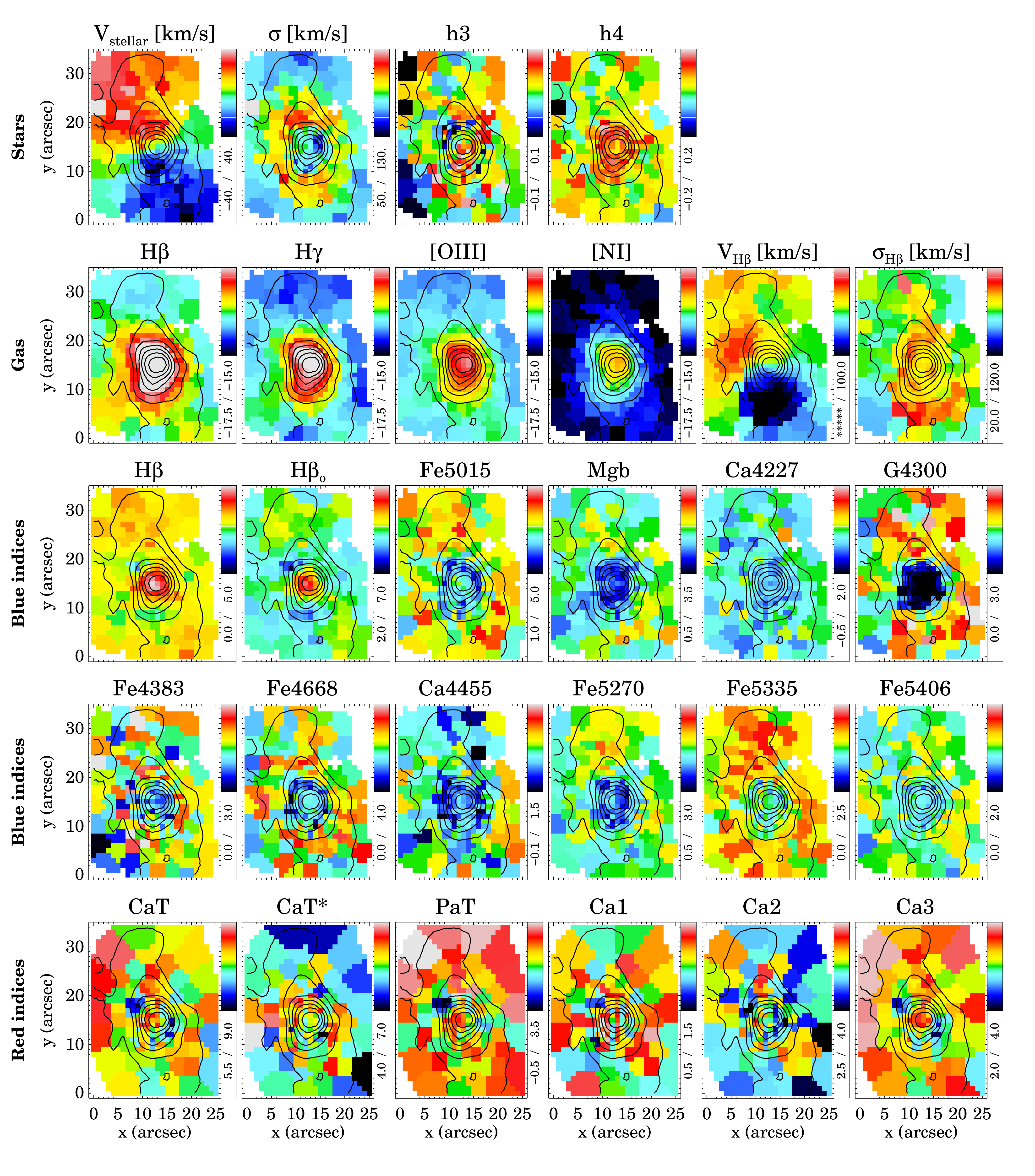}
 \caption{Summarized maps for NGC\,7552 from blue and red gratings. See text for details. }
 \label{7552all}
 \end{figure*}
%-----------------------------------------------------------------------------

\label{lastpage}

\end{document}